\documentclass[aps,prl,twocolumn,amsmath,amssymb,nofootinbib,superscriptaddress]{revtex4-1}
\usepackage{times}
\usepackage[pdftex]{graphicx}
\usepackage{dcolumn}
\usepackage{bm}
\usepackage{amsmath}
\usepackage{indentfirst}
\usepackage{float}
\usepackage[colorlinks]{hyperref}
\usepackage[dvipsnames]{xcolor}
\usepackage{xcolor}
\colorlet{shadecolor}{gray!40}

\usepackage{subfigure}
\usepackage{algorithm}
\usepackage{algpseudocode}
\usepackage{amsmath}
\usepackage{verbatim}
\usepackage{makecell}
\usepackage[normalem]{ulem}
\usepackage{setspace}
\usepackage{physics}

\begin{document}

\title{Low-Overhead Defect-Adaptive Surface Code with Bandage-Like Super-Stabilizers}

\author{Zuolin Wei}
\affiliation{Hefei National Research Center for Physical Sciences at the 
Microscale and School of Physical Sciences, University of Science and 
Technology of China, Hefei 230026, China}
\affiliation{Shanghai Research Center for Quantum Science and CAS Center for 
Excellence in Quantum Information and Quantum Physics, University of Science 
and Technology of China, Shanghai 201315, China}

\author{Tan He}
\author{Yangsen Ye}
\author{Dachao Wu}
\author{Yiming Zhang}
\author{Youwei Zhao}
\author{Weiping Lin}
\affiliation{Hefei National Research Center for Physical Sciences at the Microscale and School of Physical Sciences, University of Science and Technology of China, Hefei 230026, China}
\affiliation{Shanghai Research Center for Quantum Science and CAS Center for Excellence in Quantum Information and Quantum Physics, University of Science and Technology of China, Shanghai 201315, China}

\author{He-Liang Huang}
\email{quanhhl@ustc.edu.cn}
\affiliation{Henan Key Laboratory of Quantum Information and Cryptography, Zhengzhou, Henan 450000, China}

\author{Xiaobo Zhu}
\email{xbzhu16@ustc.edu.cn}
\author{Jian-Wei Pan}
\email{pan@ustc.edu.cn}
\affiliation{Hefei National Research Center for Physical Sciences at the Microscale and School of Physical Sciences, University of Science and Technology of China, Hefei 230026, China}
\affiliation{Shanghai Research Center for Quantum Science and CAS Center for Excellence in Quantum Information and Quantum Physics, University of Science and Technology of China, Shanghai 201315, China}
\affiliation{Hefei National Laboratory, University of Science and Technology of China, Hefei 230088, China}



\begin{abstract}
To make practical quantum algorithms work, large-scale quantum processors protected by error-correcting codes are required to resist noise and ensure reliable computational outcomes. However, a major challenge arises from defects in processor fabrication, as well as occasional losses or cosmic rays during the computing process, all of which can lead to qubit malfunctions and disrupt error-correcting codes' normal operations. In this context, we introduce an automatic adapter to implement the surface code on defective lattices. Unlike previous approaches, this adapter leverages newly proposed bandage-like super-stabilizers to save more qubits when defects are clustered, thus enhancing the code distance and reducing super-stabilizer weight. For instance, in comparison with earlier methods, with a code size of 27 and a random defect rate of 2\%, the disabled qubits decrease by $1/3$, and the average preserved code distance increases by 63\%. This demonstrates a significant reduction in overhead when handling defects using our approach, and this advantage amplifies with increasing processor size and defect rates. Our work presents a low-overhead, automated solution to the challenge of adapting the surface code to defects, an essential step towards scaling up the construction of large-scale quantum computers for practical applications.
\end{abstract}

\maketitle

\section{Introduction}

Overcoming the challenges posed by realistic hardware noise, quantum error correction (QEC) plays a pivotal role in protecting fragile qubits from decoherence effects, unlocking quantum computing's full potential. Among various QEC codes, the surface code~\cite{kitaev2003fault, fowler2012surface,raussendorf2007fault} stands out due to its 2D nearest-neighbor coupling lattice and high error threshold, typically around 1\%. While significant progress has been made in small-scale implementations of the surface code, such as increasing the code distance from two~\cite{andersen2020repeated,google2021exponential,marques2022logical} to three~\cite{zhao2022realization,krinner2022realizing} and then to five~\cite{google2023suppressing}, as well as demonstrating logical gates~\cite{bluvstein2024logical,erhard2021entangling}, the road to achieving large-scale, practical algorithms demands thousands of logical qubits with extremely low gate error rates, typically below $10^{-10}$. This necessitates millions of physical qubits~\cite{gidney2021factor}, a scale far beyond the capabilities of current physical devices.

As we strive to build larger quantum devices with more qubits, defects like nonfunctional qubits or failed entangling gates during fabrication are unavoidable. It is estimated that approximately 2\% of the qubits on a transmon device would be defective with current technology~\cite{smith2022scaling}. Even advanced processors like \textit{Zuchongzhi}~\cite{wu2021strong,zhu2022quantum,zhao2022realization,ye2023logical,gong2023quantum} and Google's Sycamore~\cite{arute2019quantum, google2023suppressing}, with just a few dozen qubits, are susceptible to defects. Additionally, external events such as cosmic rays impacting superconducting devices~\cite{martinis2021saving,mcewen2022resolving,wilen2021correlated,vepsalainen2020impact,cardani2021reducing}, or leakage and loss events in ion trap or neutral atom arrays~\cite{vala2005quantum, bermudez2017assessing, cong2022hardware} can mimic defects. Topological codes rely on specific lattice structures to encode logical states, making them susceptible to defect errors that alter the topology and reduce code distance, necessitating an adaptive approach for error correction on defective lattices~\cite{stace2009thresholds, nagayama2017surface, PhysRevA.96.042316, PhysRevApplied.19.064081, Siegel2023adaptivesurfacecode, lin2024codesign}. To this end, we introduce an adapter that deforms defective lattices and identifies super-stabilizers, enabling the implementation of the surface code on defective lattices. This adapter automates the entire process, which is crucial for scalability. As chip sizes grow, manually designing the adapter based on processor defects becomes challenging, especially since we aim for programmable logical operations. Additionally, our approach introduces a new type of super-stabilizer called bandage-like super-stabilizers. These super-stabilizers ensure that our adapter operates with low overhead when dealing with defects. Compared to previous methods~\cite{PhysRevA.96.042316, PhysRevApplied.19.064081, Siegel2023adaptivesurfacecode, lin2024codesign}, our approach minimizes the number of disabled qubits caused by defects as possible, achieving higher code distances and lower-weight super-stabilizers, thus significantly reducing logical error rates. These advantages highlight that our low-overhead defect-adaptive surface code approach provides a reliable and efficient path for scalable, large-scale fault-tolerant quantum computing.

\section{Defective Lattice Surface Code Adapter}

\begin{figure*}[!ht]
    \centering
    \includegraphics[width=1\linewidth]{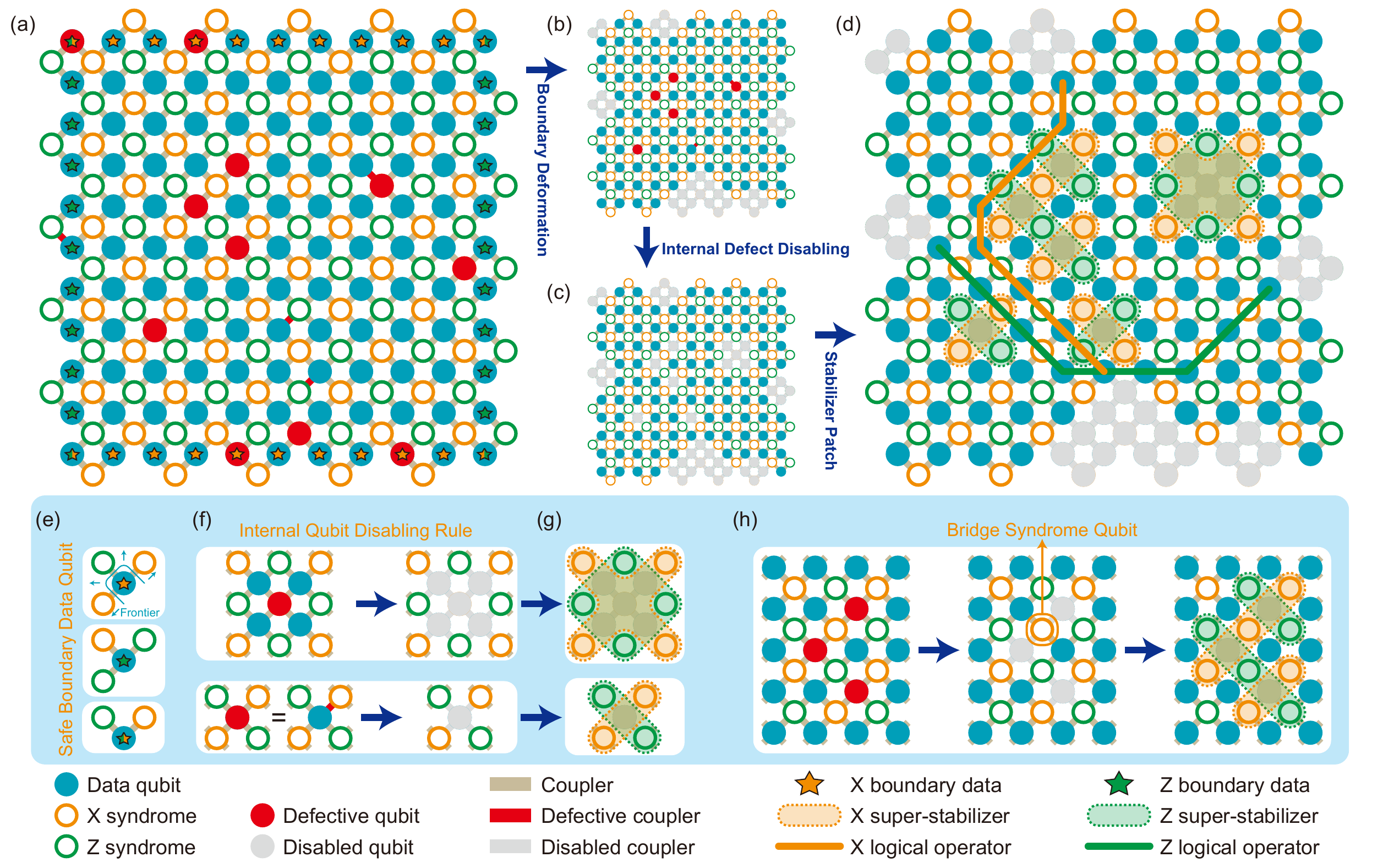}
    \caption{\textbf{The Construction Steps for the Defect-Adaptive Surface Code.} (a) An example of a defective surface code lattice, where defective qubits and couplers are marked in red, and boundary qubits are marked with star symbols. (b), (c), and (d) display the surface code lattice after undergoing boundary deformation, internal defect disabling, and stabilizer patch, respectively. (e) Illustrates safe boundary data qubits and their frontiers, including $X$, $Z$, and corner boundary data qubits from top to bottom, along with their frontiers, encompassing couplers and syndrome qubits around data qubits. (f) Depicts the rules for internal defect disabling, showcasing the disabled qubits rules for defect syndrome qubits, data qubits, and couplers from top to bottom. The rules for defect data qubits and couplers are the same. (g) and (h) demonstrate the rules for bandage-like super-stabilizers. In scenarios where internal defect qubits are not clustered, they behave similarly to traditional super-stabilizers, as shown in (g). However, in clustered situations, these super-stabilizers can stretch across weight-1 and bridge syndrome qubits, as illustrated in (h). Additionally, (h) highlights a bridge syndrome qubit for illustration purposes.}
    \label{fig:DLSCA}
\end{figure*}

Creating a surface code adapter for defective lattices demands an automated solution capable of handling diverse defect scenarios that manifest randomly across the lattice, whether along its edges or clustered closely together. Additionally, we aim to retain as many qubits as possible to mitigate the loss of error-correction capability caused by defects. In pursuit of this objective, we present a fully automated adapter customized for the surface code on a defective lattice, as depicted in Fig.~\ref{fig:DLSCA}. This adapter comprises three sequential subroutines:

\textbf{Boundary Deformation:} We kick off by addressing defects along the boundary, removing unsafe boundary data qubits and redundant syndrome qubits. A boundary data qubit is marked as safe if it ticks off three conditions: the qubit itself and its surrounding frontier, including neighboring undisabled syndrome qubits and couplers, are defect-free, and its surrounding frontier aligns with the boundary type, as shown in Fig.~\ref{fig:DLSCA}(e). This requirement stems from the surface code's need for specific syndromes to catch certain errors—an $X$($Z$) boundary data qubit, for instance, requires a $Z$($X$) syndrome to detect an $X$($Z$) error, and two $X$($Z$) syndromes to detect a $Z$($X$) error. Meanwhile, corner data qubits each require an $X$ and a $Z$ syndrome to catch errors in both directions.

Following these safety guidelines, we turn to a Breadth-First Search (BFS) algorithm to adjust the surface code lattice boundaries inward. This process, elaborated in the ``Boundary Deformation" algorithm in the Supplementary Material, 
assesses the safety of every boundary data qubit.
Upon detecting an unsafe data qubit, we disable it along with its surrounding redundant syndrome qubits, which may be defective, weight-0 (a weight-$n$ syndrome qubit has $n$ undisabled data qubit neighbors), or of different types from the boundary (e.g., $Z$ syndrome qubits adjacent to the disabled $X$ boundary data qubit).
The BFS iteratively reassesses boundary data qubits to identify any new unsafe ones and redundant syndrome qubits until no new unsafe boundary data qubits emerge, resulting in a defect-free boundary, as depicted in Fig.~\ref{fig:DLSCA}(b).

\textbf{Internal Defect Disabling:}
This straightforward step involves tackling internal defects, which come in three types: data qubit defects, syndrome qubit defects, and coupler defects.
We follow the rules outlined in Fig.~\ref{fig:DLSCA}(f) to disable these defects and their neighboring qubits. Specifically, for data qubit and coupler defects, we disable the corresponding data qubits. For syndrome qubit defects, we disable the corresponding syndrome qubits and their neighboring data qubits. The underlying reason for these rules is that internal data qubits require two $X$ and two $Z$ syndrome qubits to detect $Z$ and $X$ errors.
A specific order is necessary when tackling internal defects—defective syndrome qubit first, then defective data qubit, and finally defective coupler—to ensure each rule is applied only once and prevent conflicts.
Finally we disable weight-0 syndrome qubits caused by the implementation of the above rules. As can be easily seen, this entire process is conducted without altering the boundary's shape.

We note that internal defects may cluster together, especially at high defect rates. This leads to two primary scenarios: weight-1 syndrome qubit, and bridge syndrome qubit (see Fig.~\ref{fig:DLSCA}(h)), where a syndrome qubit connects to two active data qubits along the same diagonal line. Previous research~\cite{lin2024codesign} suggests disabling these types of syndrome qubits. However, such action may require reapplying internal defect disabling rules, potentially disrupting the previously fixed boundary shape. In cases of high defect rates, this could trigger an avalanche effect, disabling a significant portion of qubits (refer to the Supplemental Material for an example). In our approach, we don't disable internal weight-1 and bridge syndrome qubits due to our proposed bandage-like super-stabilizer. This strategy helps reduce disabled qubits, minimizing super-stabilizer weight and preventing an avalanche effect. 
Additionally, retaining bridge syndrome qubits potentially maintains a greater code distance (refer to the Supplemental Material for an example).

\textbf{Stabilizer Patch:} In this step, we utilize the proposed bandage-like super-stabilizers, which combine the same type of gauge syndrome qubits through disabled qubits, to cover all internal disabled qubits. When internal defect qubits are not clustered, they function similarly to traditional super-stabilizers~\cite{PhysRevA.96.042316, PhysRevApplied.19.064081, Siegel2023adaptivesurfacecode, lin2024codesign}, as depicted in Fig.~\ref{fig:DLSCA}(g). However, if internal defect qubits cluster, these super-stabilizers can stretch across weight-1 and bridge syndrome qubits, maintaining the integrity of syndrome qubits and conserving more data qubits, as shown in an example in Fig.~\ref{fig:DLSCA}(h). These super-stabilizers share an even number of data qubits with the opposite type of stabilizers, allowing them to commute with each other. 

In the final step, we place logical operators $X$ and $Z$, onto the defective lattice.
These operators are positioned along paths containing opposing types of syndrome qubits to avoid intersecting super-stabilizers and introducing gauge qubits.
Generally, multiple equivalent logical operators exist, and we choose the most convenient option.
Finally, our method adapts the defect lattice depicted in Fig.~\ref{fig:DLSCA}(a) into surface code shown in Fig.~\ref{fig:DLSCA}(d), resulting in a greater $X$ distance of 5 compared to the 4 achieved by the traditional method. (refer to the Supplemental Material for the performance comparison between the bandage-like method and traditional method for this defect lattice).

\section{Building Stabilizer Measurement Circuit}

\begin{figure}[!tbp]
    \centering
    \includegraphics[width=\linewidth]{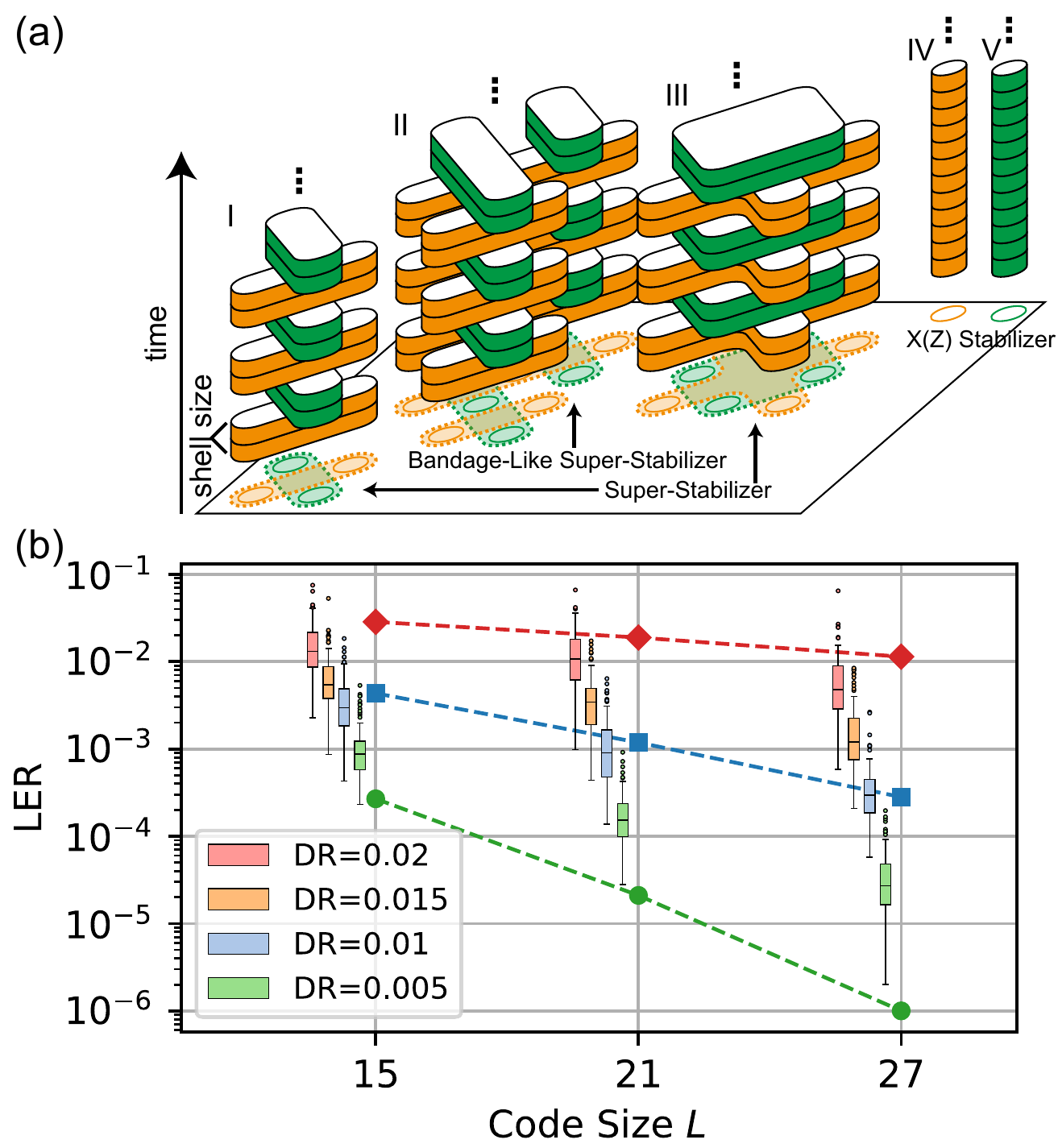}
    \caption{\textbf{Illustration of the Stabilizer Measurement Circuit Building and Related Simulation Results.} (a) Space-time lattice with super-stabilizers forming shells. The columns align along the temporal direction to show the measurement of super-stabilizers in the space-time lattice. Various types of super-stabilizers are shown: I. Super-stabilizers formed by single data qubit defect. II. Bandage-like super-stabilizers formed by nearby defect. III. Super-stabilizers formed by single syndrome qubit defect. IV. $X$ stabilizer unaffected by defects. V. $Z$ stabilizer unaffected by defects. The shell size indicates the consecutive measurements of the same type of super-stabilizers. For regular stabilizers, $X$ and $Z$ stabilizers are measured in the same cycle. However, $X$ and $Z$ super-stabilizers cannot be measured in the same cycle. (b) The Logical error rate (LER) of the surface code under different code sizes \textit{L} and defect rates (DR). The box plot displays the logical error rates for defect rates of 0.005, 0.01, 0.015, and 0.02 at a physical error rate of $p=0.002$.
    The whiskers extend to data within 1.5 times the IQR from the box. Points beyond are fliers.
    The green, blue, and red dashed lines represent references for a perfect surface code with physical error rates of $p=0.002$, $0.003$, and $0.004$, respectively. In simulations, we generated 100 devices with randomly distributed defects for each $L$ and defect rate.}
    \label{fig:bc}
\end{figure}

Building stabilizer measurement circuits for adapted devices involves measuring super-stabilizers, which can't be directly measured like regular single-syndrome stabilizers because they contain anti-commuting gauge operators.
We measure super-stabilizers using a common method from Ref.~\cite{PhysRevA.96.042316}, where $X$ and $Z$ super-stabilizers are measured in alternate cycles, and their outcomes are inferred from the gauge operators' product.
We note that in our method, multiple bandage-like super-stabilizers may intertwine to form a super-stabilizers group (e.g., in Fig.~\ref{fig:bc}(a) II, we see a group with 2 $X$ and 2 $Z$ bandage-like super-stabilizers).
It's crucial to ensure that $X$ and $Z$ super-stabilizers in the same group are not measured in the same cycle while the same type in a group are measured simultaneously.

Furthermore, to improve error correction ability, we can use the shell method outlined in Ref.~\cite{PhysRevApplied.19.064081}. This involves repeating the measurement of the same type of gauge operator for several consecutive cycles, allowing extracting information about each gauge operator's value. The number of consecutive measurement cycles is called the shell size, as illustrated in Fig.~\ref{fig:bc}(a). We must determine the appropriate shell size for each stabilizer group while ensuring it aligns with experimental constraints. Basically, there are two strategies for determining the shell size: the global strategy applies the same shell size to all stabilizer groups, while the local strategy assigns each stabilizer group its own shell size. The selection of the shell size depends on the characteristics of the processor and physical system, as discussed in the Supplemental Material. For simplicity, in the following numerical simulations, we use the global shell method.

Once the measurement circuit is set up, we can numerically simulate and test the performance of the defective lattice surface code adapter. In our simulations, we utilize the Stim simulator \cite{Gidney2021stimfaststabilizer} and employ the SI1000 circuit-level noise model \cite{Gidney2021faulttolerant}, which is well-suited for simulating superconducting experiments, as the error model. In Fig.~\ref{fig:bc}(b), we observe that for a physical error rate $p=0.002$, varying the defect rate from 0.005 to 0.02 (with consistent defect rates for qubits and couplers) still allows us to exponentially suppress the Logical Error Rate (LER) with an increase in code size $L$ (size $L$ device has $L\times L$ data qubits).
This demonstrates that our method maintains the error correction capability of the surface code. Another observation is that lowering the defect rate can improve error suppression ability. Furthermore, we compare our results against a perfect lattice with SI1000 $p$ ranging from $0.002$ to $0.004$. We find that for a $1\%$ defect rate with $p=0.002$, the error suppression ability of our adapter is comparable to that of defect-free devices with $p=0.003$. This indicates that even at high defect rates, our adapter can perform equivalent to a defect-free lattice with physical error rate increasing only by 0.001, highlighting its practical utility.

\begin{figure*}[!ht]
    \centering
    \includegraphics[width=\linewidth]{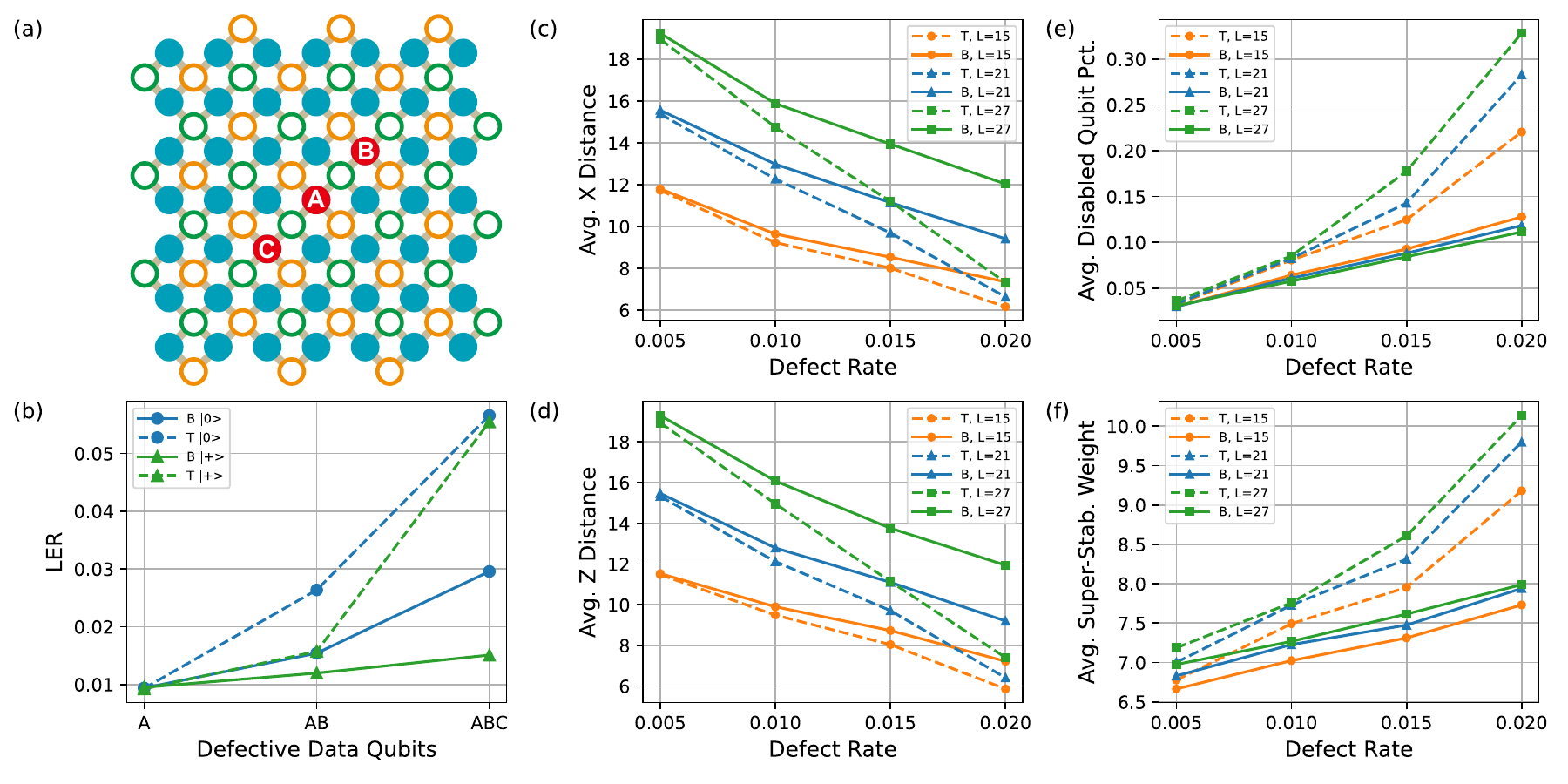}
    \caption{\textbf{Performance Comparison of the Traditional Method with Our Approach in Handling Defects.} (a) An example with a code size of $L=7$. Three highlighted circles represent potential defects, labeled as A, B, and C. We examine three scenarios: only defect A, defects A and B, and defects A, B, and C. (b) Comparison of logical error rate between bandage-like (B) and traditional (T) methods for the three scenarios in (a) at a physical error rate of $p=0.002$. The bandage-like method shows significant advantages for both $\ket{0}_L$ and $\ket{+}_L$ states. (c-f) display statistics for the bandage-like and traditional methods regarding (c) Average $X$ distance, (d) Average $Z$ distance, (e) Average disabled qubit percentage, and (f) Average super-stabilizer weight across different defect rates. Each data point is based on 100 generated devices with defects randomly distributed. In our simulations, defect rates are uniform for qubits and couplers. The bandage-like method consistently demonstrates substantial advantages, regardless of defect rates or code size. These advantages increase significantly with larger code sizes and defect rates.}
    \label{fig:bae}
\end{figure*}

\section{Bandage-Like Super-Stabilizer Advantage}

To further showcase the advantages of our approach, we compare it with traditional super-stabilizer methods~\cite{PhysRevA.96.042316, PhysRevApplied.19.064081, Siegel2023adaptivesurfacecode, lin2024codesign}.
We start with a simple case of three scenarios with increasing defects for $L=7$ devices, as seen in Fig.~\ref{fig:bae}(a).
The results are then shown in Fig.~\ref{fig:bae}(b).
When there's only one defect A, the two methods perform equally.
However, as we move from A to AB and then ABC defects, the advantage of the bandage-like method becomes more pronounced. Specifically, the bandage method lowers the LER by 42\% (24\%) for the $\ket{0}_L$ ($\ket{+}_L$) state with 2 defects, AB, and this improves to 48\% (73\%) for 3 defects, ABC. This improvement happens because the bandage-like method keeps more code distance and lower-weight stabilizers.
For the AB defect scenario, the traditional method provides a $X$ ($Z$) distance of 5 (5), while the bandage-like method maintain 5 (6). Moreover, the average super-stabilizer weight decreases from 10 to 6.67.
Similarly, in the ABC defect case, the traditional method's $X$ ($Z$) distance is 4 (4), while the bandage-like method provides 4 (6). And the average super-stabilizer weight drops from 14 to 7 (refer to Supplementary Material for details).

To generalize this advantage to a broader context, we compare our bandage-like super-stabilizer approach with traditional methods~\cite{PhysRevA.96.042316, PhysRevApplied.19.064081, Siegel2023adaptivesurfacecode, lin2024codesign} for adapting multiple devices with randomly distributed defects. It's worth noting that we intentionally kept weight-1 and bridge syndrome qubits near boundary data qubits to prevent further boundary deformation for the traditional methods. This design choice slightly favored the traditional method, resulting in a higher baseline performance. In our simulations, we generated 100 devices with randomly distributed defects for each scenario for statistical purposes, and the defect rates are consistent across qubits and couplers (refer to the Supplemental Material for scenarios of devices with only coupler defects).

Figure~\ref{fig:bae}(c-d) shows the average code distance after adaptation with increasing defect rates. Naturally, code distance decreases with higher defect rates for both methods. However, the bandage-like method consistently maintains a superior code distance, and this advantage grows as defect rates increase, similar to the above specific case. For instance, at a code size of $L=27$ and a defect rate of 0.01 (0.02), the average $X$ distance improves from 14.8 (7.3) to 15.9 (12.0), and the $Z$ distance from 15.0 (7.4) to 16.1 (11.9), marking a 7.6\%~($63\%$) average improvement.

The bandage-like method preserves code distance by disabling fewer qubits. To quantitatively illustrate this, Fig.~\ref{fig:bae}(e) shows the average percentage of disabled qubits after adaptation. While disabled qubits increase with defect rates, the bandage method exhibits a slower increase, indicating better qubit preservation compared to the traditional method. For instance, at a code size of $L=27$ and a defect rate of 0.01~(0.02), the average percentage of disabled qubits improves from 8.5\% (32.8\%) to 5.8\% (11.1\%).

Additionally, Fig.~\ref{fig:bae}(f) shows the weighted average of the average super-stabilizer weight for each random device, calculated as:
$w_\text{avg}=(\sum_d\sum_i w_{di})/(\sum_d \sum_i 1)$. Here, $d$ represents the index of the random device (100  devices with defects
randomly distributed in our simulation), and $i$ represents the index of the super-stabilizer in each device. $w_{di}$ represents the weight of the super-stabilizer indexed by $d$ and $i$. We can observe that the bandage-like method exhibits lower average super-stabilizer weights. For instance, at a code size of $L=27$ and a defect rate of 0.01~(0.02), the average super-stabilizer weight improves from 7.8~(10.1) to 7.3~(8.0), marking a 6.3\%~(21\%) improvement. This reduction in weight enhances the reliability of the super-stabilizer because lower-weight super-stabilizers are better at identifying errors within a more localized area, thus preventing error spread and enhancing error detection and correction capabilities.

Analyzing Fig.~\ref{fig:bae}(c-f) collectively, we observe another significant trend: as code size increases under the same defect rates, the bandage-like method's advantages over the traditional approach also increase in terms of  code distance, disabled qubit percentage, and  super-stabilizer weight. This scalability advantage is crucial for scaling quantum computing devices. 

\section{Conclusion and Outlook}
The proposed defect-adaptive surface code has two main features: 1) Our algorithm automates defect handling step by step, including boundary deformation, internal defect disabling, stabilizer patch, all without conflicts even in complex defect scenarios. This method allows us to obtain defect-adaptive surface codes automatically in all simulations, without manual intervention. Even with a high defect rate of 2\%, our method still shows an exponentially suppression in the logical error rate as the code distance increases, demonstrating the feasibility and excellent performance of our approach. These advantages are crucial for the large-scale expansion of quantum computing. 2) Unlike previous methods, our adapter utilizes a new type of bandage-like super-stabilizers, offering advantages in maintaining more qubits and code distance, and reducing super-stabilizer weight, especially in scenarios with clustered defects. This significantly reduces the logical error rate of quantum memory on defective lattices. In a simulation with three defects, our method reduces the logical error rate by 48\% and 73\% for the $\ket{0}_L$ and $\ket{+}_L$ states compared to previous works. Future interesting work will involve observing the performance of our experimentally-ready approach in real-world experiments and achieving error suppression by scaling surface code logical qubits under defective lattices.

\begin{acknowledgments}
The authors are grateful for valuable discussions with Armands Strikis and  Ke Liu. 
\textbf{Funding:}
This research was supported by the Chinese Academy of Sciences, Anhui 
Initiative in Quantum Information Technologies, Shanghai Municipal Science and 
Technology Major Project (Grant No. 2019SHZDZX01), Innovation Program for 
Quantum Science and Technology (Grant No. 2021ZD0300200), Special funds from 
Jinan science and Technology Bureau and Jinan high tech Zone Management 
Committee, Technology Committee of Shanghai Municipality, National Science 
Foundation of China (Grants No. 11905217, No. 11774326), Natural Science Foundation of Shandong Province, China (grant number
ZR202209080019), and Natural Science Foundation of Shanghai (Grant No. 23ZR1469600), the Shanghai Sailing Program (Grant No. 23YF1452600). X. B. Zhu acknowledges support 
from the New Cornerstone Science Foundation through the XPLORER PRIZE. H.-L. H. acknowledges support from the National Natural Science Foundation of China (Grant No. 12274464), Natural Science Foundation of Henan (Grant No. 242300421049), and Youth Talent Lifting Project (Grant No. 2020-JCJQ-QT-030).

\end{acknowledgments}

\bibliographystyle{apsrev4-1}
\bibliography{references}

\end{document}


\title{Supplemental Material for ``Low-Overhead Defect-Adaptive Surface Code with Bandage-Like Super-Stabilizers"}

\author{Zuolin Wei}
\affiliation{Hefei National Research Center for Physical Sciences at the 
Microscale and School of Physical Sciences, University of Science and 
Technology of China, Hefei 230026, China}
\affiliation{Shanghai Research Center for Quantum Science and CAS Center for 
Excellence in Quantum Information and Quantum Physics, University of Science 
and Technology of China, Shanghai 201315, China}

\author{Tan He}
\author{Yangsen Ye}
\author{Dachao Wu}
\author{Yiming Zhang}
\author{Youwei Zhao}
\author{Weiping Lin}
\affiliation{Hefei National Research Center for Physical Sciences at the Microscale and School of Physical Sciences, University of Science and Technology of China, Hefei 230026, China}
\affiliation{Shanghai Research Center for Quantum Science and CAS Center for Excellence in Quantum Information and Quantum Physics, University of Science and Technology of China, Shanghai 201315, China}

\author{He-Liang Huang}
\email{quanhhl@ustc.edu.cn}
\affiliation{Henan Key Laboratory of Quantum Information and Cryptography, Zhengzhou, Henan 450000, China}
\author{Xiaobo Zhu}
\email{xbzhu16@ustc.edu.cn}
\author{Jian-Wei Pan}
\email{pan@ustc.edu.cn}
\affiliation{Hefei National Research Center for Physical Sciences at the Microscale and School of Physical Sciences, University of Science and Technology of China, Hefei 230026, China}
\affiliation{Shanghai Research Center for Quantum Science and CAS Center for Excellence in Quantum Information and Quantum Physics, University of Science and Technology of China, Shanghai 201315, China}
\affiliation{Hefei National Laboratory, University of Science and Technology of China, Hefei 230088, China}


\pacs{03.65.Ud, 03.67.Mn, 42.50.Dv, 42.50.Xa}

\maketitle

\setcounter{section}{0}
\renewcommand{\thefigure}{S\arabic{figure}}	
\renewcommand{\thetable}{S\arabic{table}}	
\renewcommand{\theequation}{S\arabic{equation}}	
\setcounter{figure}{0}
\setcounter{table}{0}
\setcounter{equation}{0}

\section{Defective lattice surface code Adapter}
The adapter we propose achieves the creation of functional surface codes on defective lattices through three key steps: boundary deformation, internal defect disabling, and stabilizer patching. Let's explore each of these steps and algorithms thoroughly. In the upcoming algorithms, we'll approach it from a graphical perspective, treating qubits (whether data or syndrome qubits) and couplers within the lattice as nodes and edges on a graph, respectively.

\subsection{Boundary Deformation}

First up is boundary deformation, aimed at handling defects along the boundary. We'll kick things off by laying out the definitions for the different types of boundary data nodes.

\begin{figure}[htbp]
    \centering
    \includegraphics[width=0.7\linewidth]{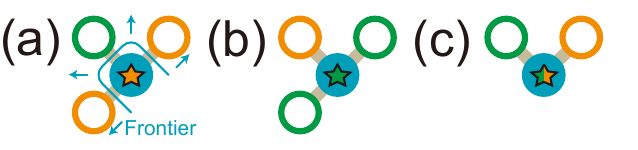}
    \caption{\textbf{Safe boundary data nodes and theirs frontiers.} The frontiers of a boundary data node comprise neighboring syndrome nodes along with the connecting edges. A boundary data node is safe if the node itself and its frontier are defect-free, and its frontier should match the type of boundary data node: 
    (a) An $X$-type boundary data node should have a frontier containing 2 $X$ syndrome nodes and 1 $Z$ syndrome node.
    (b) A $Z$-type boundary data node should have a frontier containing 2 $Z$ syndrome nodes and 1 $X$ syndrome node.
    (c) A $C$-type boundary data node should have a frontier containing 1 $Z$ syndrome node and 1 $X$ syndrome node.
    }
    \label{fig:frontier}
\end{figure}

\begin{myDef}
    \textbf{Types of Boundary Data Node}:
    \begin{itemize}
        \item A boundary data node exclusively situated on $X$/$Z$ boundary is categorized as a $X$/$Z$-type boundary data node (see Fig.~\ref{fig:frontier}(a) and Fig.~\ref{fig:frontier}(b)).
        \item A boundary data node located on both $X$ and $Z$ boundaries is classified as a $C$-type (Corner) boundary data node (see Fig.~\ref{fig:frontier}(c)).
    \end{itemize}
\end{myDef}

\begin{figure*}[htbp]
    \centering
    \includegraphics[width=0.8\linewidth]{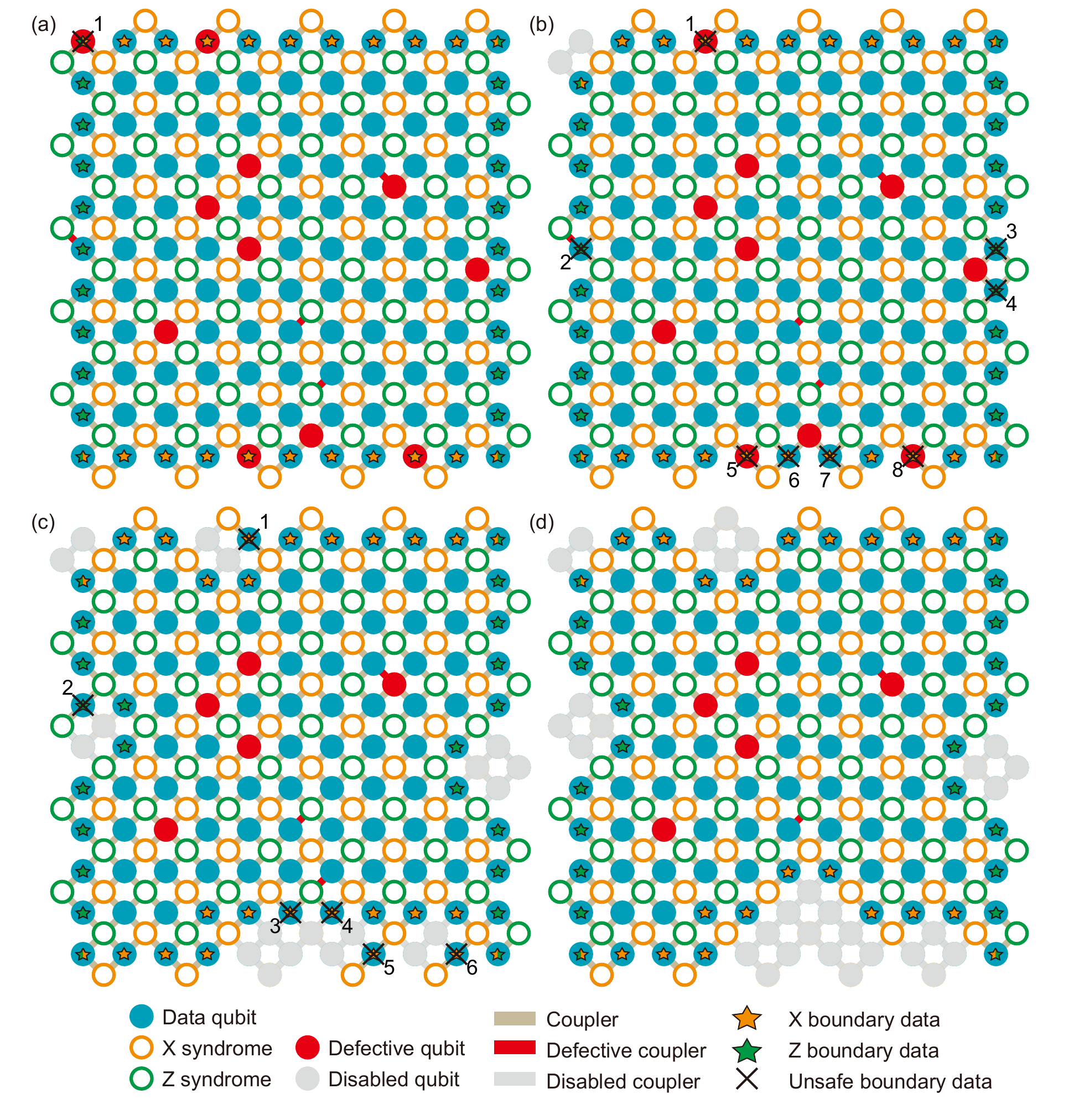}
    \caption{\textbf{
        Detailed steps for the boundary deformation of defective lattice in the Fig.~1(a) of the main text.} The entire process is illustrated sequentially from (a) to (d). 
        (a) Initial boundary data nodes are highlighted with colored stars. The top and bottom boundaries correspond to $X$ boundaries, while the left and right boundaries correspond to $Z$ boundaries. Intersections of $X$ and $Z$ boundaries are denoted as $C$ (corner) boundary data nodes. Step 1 addresses corner boundary data qubits first. As shown, one corner boundary data node is unsafe due to being defective, and should be disabled. Further, the $Z$ syndrome neighbors to it should be cleaned according to the frontier cleaner subroutine. (b) Step 2 is introduced to address the remaining boundary data node defects. Several scenarios are encountered here: Unsafe boundary data nodes 1, 5 and 8 are defective themselves; unsafe node 2 has defective edge in the frontier; unsafe nodes 3, 4, 6 and 7 have defective syndrome node in the frontier. All these unsafe boundary data nodes need to be disabled, and the syndrome nodes neighboring the unsafe ones should be cleaned using the frontier cleaner, including isolated syndrome nodes and syndrome nodes with the wrong type from the boundary. (c) Step 3 is similar to step 2. Unsafe boundary data nodes 1-6 have an incorrect frontier shape with its boundary type. (d) Upon completing the above process, a clean boundary with all boundary data nodes safe is obtained.
    }
    \label{fig:bdstep}
\end{figure*}

Figure~\ref{fig:bdstep}(a) provides an example, marking the different types of boundary data nodes in a surface code. Following that, we define the frontier of a boundary data node as the set of its neighboring syndrome nodes and the edges connecting to these nodes, as illustrated in Fig.~\ref{fig:frontier}. With these concepts, we can determine if a boundary data node is safe, which must meet the following three conditions simultaneously:

\begin{itemize}
    \item \textbf{\textit{Condition-1}}: The boundary data node itself is defect-free;
    \item \textbf{\textit{Condition-2}}: The components in its frontier, including syndrome nodes and edges, are defect-free;
    \item \textbf{\textit{Condition-3}}: The type and number of syndrome qubits on the frontier should match the type of the boundary data node, as illustrated in Fig.~\ref{fig:frontier}.
\end{itemize}

\begin{figure}[H]
\begin{algorithm}[H]
    \caption{Boundary Deformation}
    \label{algorithm:bd}
    \begin{algorithmic}
        \State \textbf{Input:} Initial boundary data nodes
        \State \textbf{Output:} Disabled boundary nodes
        \For{$n_0$ \textbf{in} $C$ boundary data nodes}
            \If{$n_0$ is unsafe}\Comment{Visiting the corner first is better}
                    \State Disable $n_0$
                    \State \Call{FrontierCleaner}{$n_0$}
            \EndIf\Comment{Refer to Algorithm~\ref{algorithm:fcs}}
        \EndFor
        \State flag$\gets$ True
        \While{flag}
            \State flag$\gets$ False
            \For{$n_0$ \textbf{in} undisabled boundary data nodes}
                \If{$n_0$ is unsafe}
                    \State flag$\gets$ True
                    \State Disable $n_0$
                    \State \Call{FrontierCleaner}{$n_0$}
                \EndIf\Comment{Refer to Algorithm~\ref{algorithm:fcs}}
            \EndFor
        \EndWhile
    \end{algorithmic}
\end{algorithm}
\end{figure}

We remove all the unsafe boundary data nodes using the boundary deformation algorithm, detailed in Algorithm~\ref{algorithm:bd}. This iterative algorithm evaluates the safety of each boundary data node, starting with corner boundary data nodes in the initial round and then assessing all boundary data nodes in subsequent rounds. If an unsafe data node is detected, it needs to be disabled, and then the frontier cleaner described in Algorithm~\ref{algorithm:fcs} is called to clean its frontier. This involves removing weight-0 syndrome nodes, defective syndromes, and syndrome qubits of different types from the boundary, such as the $Z$ syndrome qubit adjacent to the disabled $X$ boundary data qubit. After implementing these, new boundary data nodes are introduced, pushing the boundary inward. This iterative process continues until no unsafe boundary data nodes remain within the boundary. Figure~\ref{fig:bdstep} illustrates a sample of boundary deformation steps for the defect lattice in Fig.~1(a) of the main text.

\subsection{Internal Defect Disabling}

\begin{figure}[!h]
    \centering
    \includegraphics[width=0.6\linewidth]{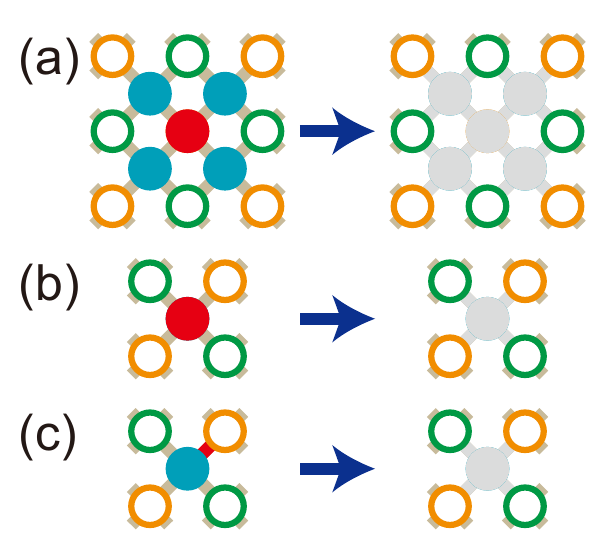}
    \caption{\textbf{Internal defect disabling rules.} (a) Rule for a defective syndrome node; (b) Rule for a defective data node; (c) Rule for a defective coupler. Red indicates defective components, while gray represents disabled components.}
    \label{fig:idh}
\end{figure}

The step of \textit{Internal Defect Disabling} addresses internal defects. This process is straightforward; by applying the following rules (see Fig.~\ref{fig:idh}) just once, internal defects and their related qubits can be effectively removed.

\begin{itemize}
    \item \textit{\textbf{Step 1:}} Disable the defective syndrome nodes and their neighboring data nodes, as shown in Fig.~\ref{fig:idh}(a).
    \item \textit{\textbf{Step 2:}} Disable all defective data nodes, as illustrated in Fig.~\ref{fig:idh}(b).
    \item \textit{\textbf{Step 3:}} Disable data nodes if the edges (couplers) connected to them are defective, as depicted in Fig.~\ref{fig:idh}(c).
    \item \textit{\textbf{Step 4:}} Disable any weight-0 syndrome nodes resulting from the above steps.
\end{itemize} 

After executing these steps, all defect nodes and edges in the surface code lattice will be successfully processed. For example, in the lattice shown in Fig.~\ref{fig:bdstep}(d), after the step of \textit{Internal Defect Disabling}, we obtain the lattice shown in Fig.~\ref{fig:sg}(a).

\begin{figure*}[h]
\begin{minipage}{\linewidth}
\begin{algorithm}[H]
    \caption{Frontier Cleaner Subroutine}
    \label{algorithm:fcs}
    \begin{algorithmic}
        \State \textbf{Input:} $n_0\gets$ Disabled data node to clean its frontier
        \State \textbf{Output:} Disabled syndrome nodes in the frontier, new introduced boundary data nodes
        \State $t\gets$ boundary type of $n_0$\Comment{Type of boundary data node in BX, BZ or BC}
        \State $S_\text{f}\gets$ undisabled syndrome nodes neighbor to $n_0$
        \For{$s$ \textbf{in} $S_\text{f}$}\Comment{Disabling redundant syndrome nodes}
            \If{$s$ is defective} Disable $s$, \textbf{continue}
            \ElsIf{$s$ is weight-0 syndrome node} Disable $s$, \textbf{continue}
            \ElsIf{$s$ is $X$ syndrome node \textbf{and} $t=\text{BZ}$} 
                Disable $s$, \textbf{continue}
            \ElsIf{$s$ is $Z$ syndrome node \textbf{and} $t=\text{BX}$} 
                Disable $s$, \textbf{continue}
            \EndIf
        \EndFor
        \If{$t=\text{BC}$}
        \State $S'_\text{f}\gets$ undisabled syndrome nodes neighbor to $n_0$ \Comment{$S'_\text{f}$ differs from $S_\text{f}$}
            \If{len($S'_\text{f}$) $=2$}
                \State Disable the node with fewer undisabled neighbors; if they are equal, disable the $Z$ syndrome node.
            \EndIf\Comment{Prefer $X$ syndrome nodes}
        \EndIf
        \State $S_\text{d} \gets$ disabled syndrome nodes that have undisabled neighbors and are neighbors to $n_0$
        \If{$t=\text{BC}$}\Comment{Introducing new boundary data nodes, $C$ Boundary situation}
            \State $x\gets 0, z\gets0$
            \For{$s$ \textbf{in} $S_\text{d}$}
                \State $N_0 \gets$ undisabled nodes that have 3 undisabled neighbors and are neighbors to $s$
                    \If{$s$ is $X$ syndrome node}
                        \State $x\gets x+\text{len}(N_0)$
                    \ElsIf{$s$ is $Z$ syndrome node}
                        \State $z\gets z+\text{len}(N_0)$
                \EndIf 
            \EndFor
            \If{$x<z$} $t\gets \text{BX}$\Comment{Treat corner as $X$ boundary}
            \ElsIf{$x>z$} $t\gets \text{BZ}$\Comment{Treat corner as $Z$ boundary}
            \ElsIf{$x=z$} $t\gets \text{BX}$\Comment{Prefer $X$ syndrome nodes}
            \EndIf
        \EndIf
        \For{$s$ \textbf{in} $S_\text{d}$}
            \If{$s$ is $X$ syndrome and $t=\text{BZ}$ \textbf{or} $s$ is $Z$ syndrome and $t=\text{BX}$}
                \State $N_1 \gets$ undisabled nodes neighbor to $s$
                \If{$t=\text{BX}$}\Comment{$X$ boundary situation}
                    \State Add $N_1$ to the $X$ boundary
                \ElsIf{$t=\text{BZ}$}\Comment{$Z$ boundary situation}
                    \State Add $N_1$ to the $Z$ boundary
                \EndIf
            \EndIf
        \EndFor
    \end{algorithmic}
\end{algorithm}
\end{minipage}
\end{figure*}

\begin{figure*}
    \centering
    \includegraphics[width=\linewidth]{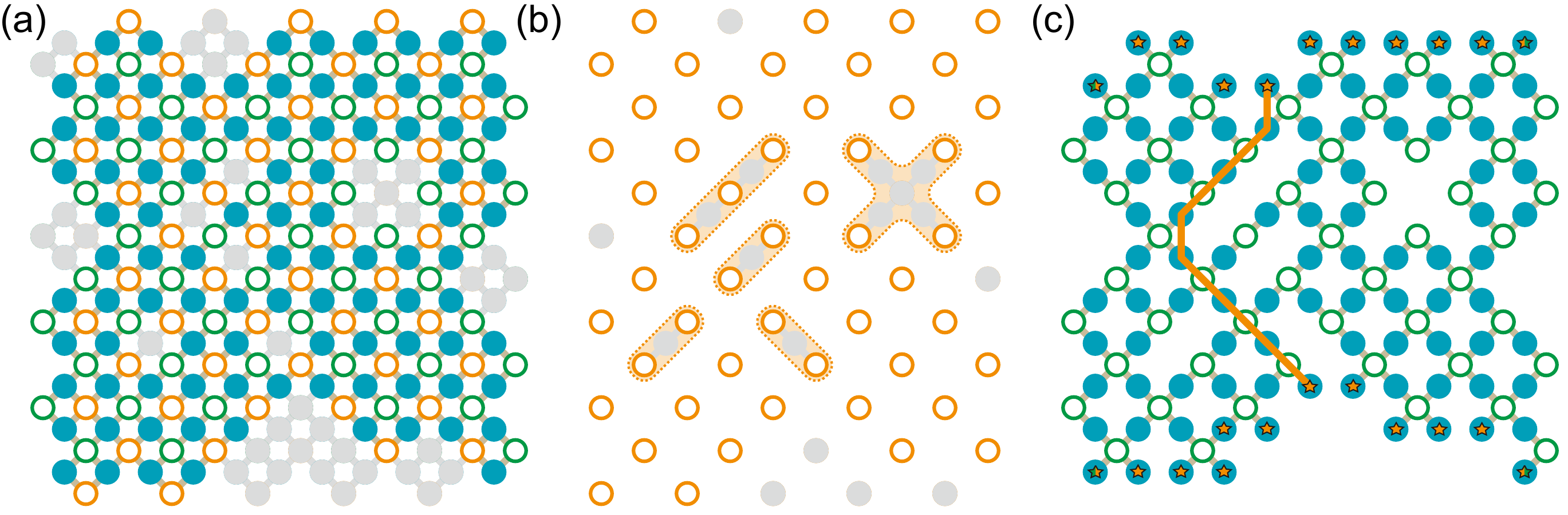}
    \caption{\textbf{Subgraphs for stabilizer and logical operator search.}
    (a) The lattice has undergone boundary deformation and internal defect disabling, rendering all defective nodes and edges (depicted in gray) inactive in this diagram. This enables us to search for  stabilizers and logical operators within its subgraphs. (b) $X$ stabilizer search graph. This subgraph is employed for searching $X$ stabilizers. It encompasses all internally deactivated data nodes, alongside all $X$ syndrome nodes and their connecting edges. Algorithm \ref{algorithm:ss} is deployed to identify all connected components on this graph containing active $X$ syndrome nodes, facilitating the creation of stabilizers, including super-stabilizers. (c) $X$ logical operator search graph. This subgraph is dedicated to the search for $X$ logical operators. It includes all active data nodes, undisabled $Z$ syndrome nodes, and the connecting edges. The shortest logical $X$ operator, highlighted by an orange line, illustrates how it can span between the top and bottom sections of the $X$ boundaries, composed of data nodes on the shortest path.
    }
    \label{fig:sg}
\end{figure*}

\subsection{Stabilizer Patching}

Since the previous steps have disabled some qubits, we need to search for super-stabilizers to ensure that the surface code can function properly. The bandage-like super-stabilizers combine the same type of gauge syndrome nodes through disabled nodes and can be searched using a path finding algorithm. For this purpose, we first define the stabilizer search graph to simplify the stabilizer search procedure by identifying connected components on it.

\begin{myDef}
    \textbf{Stabilizer Search Graph}. 

    An $X$($Z$) stabilizer search graph is a subgraph of the original surface code lattice, containing:
    \begin{itemize}
        \item All internally disabled data nodes within the lattice. 
        \item All $X$($Z$) syndrome nodes.
        \item All edges connecting the aforementioned node types.
    \end{itemize}
\end{myDef}

A sample $X$ stabilizer search graph is illustrated in Fig.~\ref{fig:sg}(b). We then utilize Algorithm~\ref{algorithm:ss} to conduct $X$($Z$) super-stabilizer search on the $X$($Z$) stabilizer search graph. This algorithm traverses all undisabled syndrome nodes on the stabilizer search subgraph and identifies all connected components. The syndrome nodes within the same connected component naturally constitute a super-stabilizer, as depicted in Fig.~\ref{fig:sg}(b). By executing the Stabilizer Search for both $X$ and $Z$ syndrome nodes, all stabilizers have now been formed, effectively addressing all defects, and they all commute.

Finally, we will discuss the placement of logical operators on a surface code to encode a logical qubit within the defective surface code lattice. To achieve this, we will introduce the concept of the logical operator search for positioning $X$/$Z$ logical operators.

\begin{myDef}
    \textbf{Logical Operator Search Graph}.

    An $X$($Z$) logical operator search graph is a subgraph of the original surface code lattice that consists of several elements:
    \begin{itemize}
        \item All undisabled data nodes
        \item All $Z$($X$) undisabled syndrome nodes
        \item All edges in the original graph that connect these elements
    \end{itemize}
\end{myDef}

A sample $X$ logical operator search graph is depicted in Fig.~\ref{fig:sg}(c). Once established, determining logical operators becomes straightforward:

\begin{itemize}
    \item The $X$ logical operator is determined by identifying the shortest path between the top and bottom boundaries of the $X$ boundaries on the $X$ logical operator search graph.
    \item Similarly, the $Z$ logical operator is determined by identifying the shortest path between the left and right boundaries of the $Z$ boundaries on the $Z$ logical operator search graph.
\end{itemize}

The selection of the shortest path is based on the requirement that, following the property of the surface code, all logical operators connecting both parts of the boundaries in the logical operator search graph must be identical after decoding.

After implementing these algorithmic steps, we have successfully adapted the surface code, achieving the following outcomes:
\begin{itemize}
    \item Nodes related to defects were disabled to preserve the properties of the stabilizer code.
    \item Establishment of stabilizers (including super-stabilizers) in the adapted code.
    \item Determination of $X$ and $Z$ logical operators intended for protection by the code.
\end{itemize}

\begin{figure*}[h]
\begin{minipage}{\linewidth}
\begin{algorithm}[H]
    \caption{Stabilizer Search}\label{algorithm:ss}
    \begin{algorithmic}
        \For{$n_s$ \textbf{in} undisabled syndrome nodes}
            \If{the $n_s$ is already part of a stabilizer}
                \State \textbf{continue}\Comment{Skip visited syndrome node.}
            \EndIf
            \State $cc \gets$ \Call{ConnectedComponent}{$n_s$}\Comment{Connected component on stabilizer search graph, as shown in Fig.~\ref{fig:sg}(b).}
            \State Combine all undisabled syndrome nodes in $cc$ to form a stabilizer.
        \EndFor
    \end{algorithmic}
\end{algorithm}
\end{minipage}
\end{figure*}

\clearpage

\section{Building Stabilizer Measurement Circuit}

\subsection{Shell Strategies}

\begin{figure}[htbp]
    \centering
    \includegraphics[width=\linewidth]{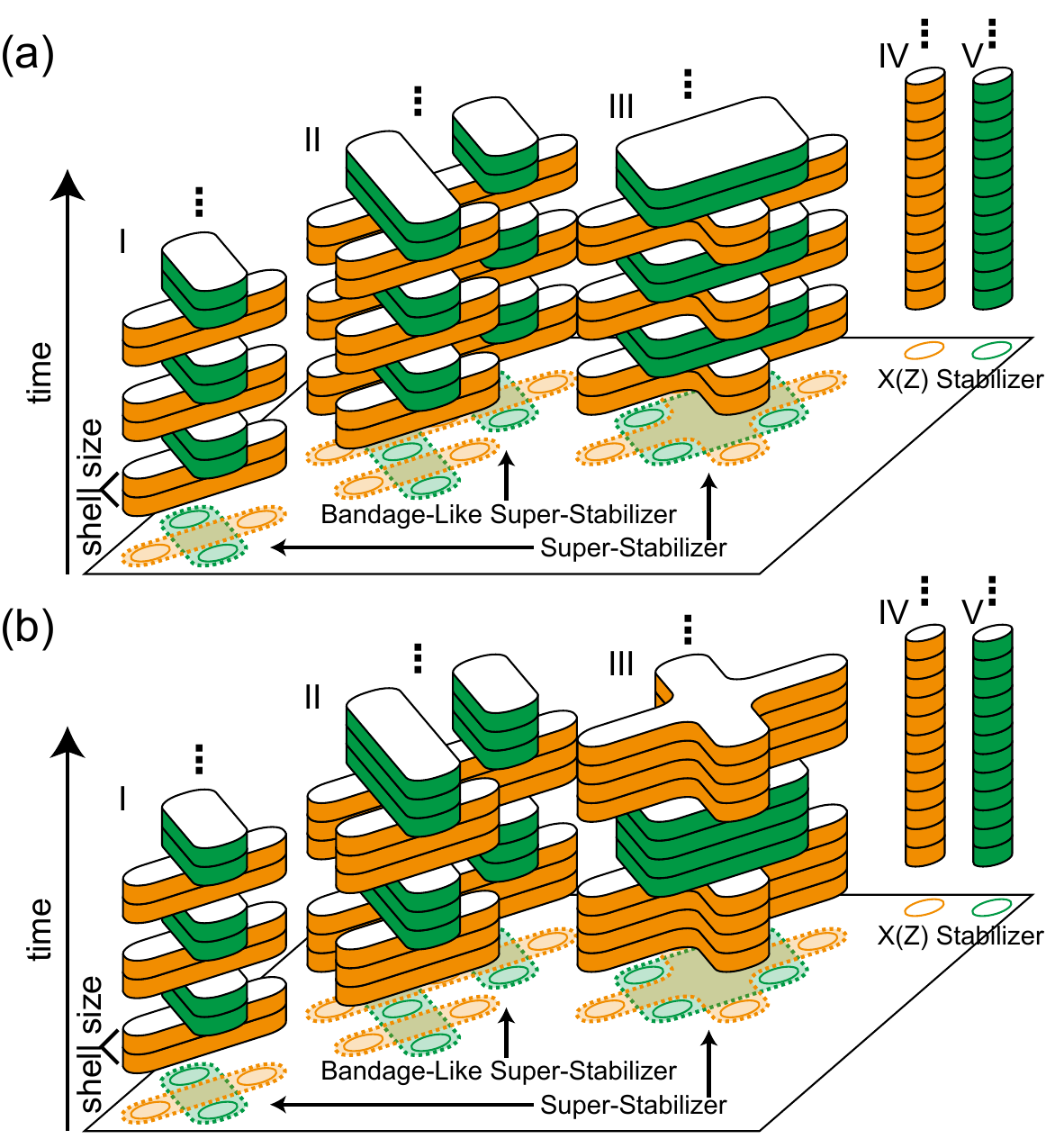}
    \caption{\textbf{Illustration of global and local shell strategies.} (a) Global strategy. The same shell size is maintained for different stabilizer groups I, II, and III. Stabilizer IV and V, unaffected by defects, are measured every cycle. (b) Local strategy. Different shell sizes are allocated for different stabilizer groups. For instance, groups I, II, and III have shell sizes of 2, 3, and 4, respectively. Stabilizers IV and V are measured every cycle, similar to the global strategy.}
    \label{fig:gals}
\end{figure}

As mentioned in the main text, when building the stabilizer circuit, we can employ two methods: using a global shell size for all stabilizer groups, or using a local shell size for each stabilizer group, as illustrated in Fig.~\ref{fig:gals}. For the local approach, further subdivision is possible. In this section, we will investigate three different shell strategies, as outlined below:

\begin{itemize}
    \item \textbf{GLOBAL}: All stabilizer groups share the same shell size $n_\text{shell}$, ranging from $1$ to $(L-1)/2$, where $L$ is the code size of the original lattice with $L\times L$ data qubits.
    \item \textbf{LOCALAVG}: Each stabilizer group has a shell size of $\lfloor r \cdot w_{\text{avg}} \rfloor$, where $r$ is the shell size-weight ratio, and $w_{\text{avg}}$ represents the average weight of super-stabilizers in each stabilizer group.
    \item \textbf{LOCALMAX}: Each stabilizer group has a shell size of $\lfloor r \cdot w_{\text{max}} \rfloor$, where $w_{\text{max}}$ represents the maximum weight of super-stabilizers in each stabilizer group.
\end{itemize}

Next, we'll demonstrate the performance of different shell size strategies from a simulation perspective. It's worth noting that in our simulations, we did not consider complex noise models like crosstalk or leakage. Therefore, considering the specific physical implementation, we recommend initially comparing these strategies through experiments to choose the one that offers better performance. And our analysis in this section can provide insights for how to select the optimal strategy.

\subsection{Settings for the Simulation}

We generate a stabilizer measurement circuit and leverage the Stim simulator \cite{Gidney2021stimfaststabilizer} for the simulation.
Our choice of the SI1000 circuit-level noise model, as described in \cite{Gidney2021faulttolerant}, is particularly suitable for simulating superconducting experiments.
The error rate for each quantum gate is outlined in Table~\ref{tab:si1000}.
We conducted simulations on 100 devices with randomly introduced defects for each defect rate and code size, and these devices maintain a consistent defect rate for both qubits and couplers (applicable to the scenario of superconducting quantum devices with tunable coupler architecture).
Logical error rates for the $\ket{0}$ and $\ket{+}$ states are simulated after $L$ cycles of stabilizer measurements, where $L$ represents the code size of the unadapted lattice.
The circuit begins by measuring the opposite super-stabilizer of the initial state to cover all types of errors that may occur at the beginning of the circuit. For instance, when considering the $\ket{0}$ state, we begin by measuring $X$ super-stabilizers.

\begin{table}[ht]
    \centering
    \begin{tabular}{|c|c|c|}
        \hline
        Gate&Error Rate&Noise Channel\\\hline\hline
        CZ & $p$ & 2-qubit depolarizing\\\hline
        AnyClifford$_1$&$p/10$&1-qubit depolarizing\\\hline
        Init$_Z$&$2p$&bitflip\\\hline
        M$_Z$&$5p$&bitflip\\\hline
        Idle&$p/10$&1-qubit depolarizing\\\hline
        ResonatorIdle&$2p$&1-qubit depolarizing\\\hline
    \end{tabular}
    \caption{\textbf{SI1000 circuit-level noise model \cite{Gidney2021faulttolerant}.} We use $p$ to indicate the physical error rate.}
    \label{tab:si1000}
\end{table}

\begin{figure}[h]
    \centering
    \includegraphics[width=\linewidth]{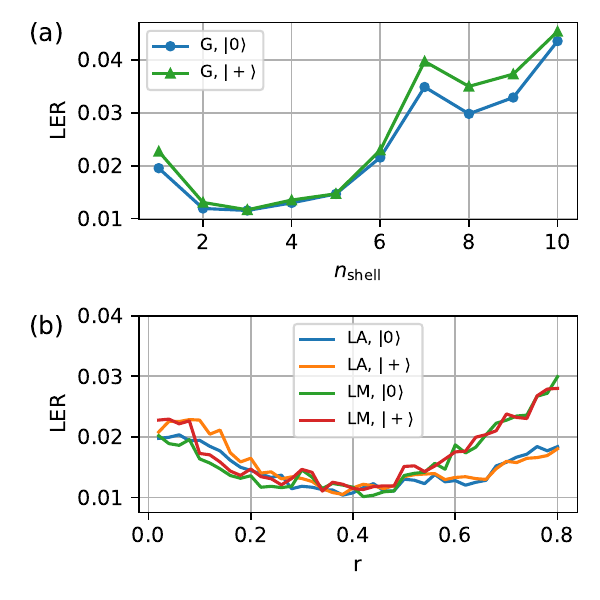}
    \caption{\textbf{Determining the optimal shell size by investigating different values of $n_\text{shell}$ for the GLOBAL(G) method, and $r$ for the LOCALAVG(LA) and LOCALMAX(LM) methods.} (a) LER versus global shell size $n_\text{shell}$ for a specific device, with a defect rate of 0.02 for qubits and couplers, code size $L=21$ and a physical error rate of $p=0.002$. (b) LER versus local shell ratio $r$ for the same specific device. Both global and local methods exhibit a trend of decreasing LER followed by an increase. The point with the lowest LER is referred to as the ``sweet point".
    }
    \label{fig:sweet}
\end{figure}

\subsection{Simulation Results}

For each strategy, we must first determine the optimal shell size. To achieve this, we analyze the trend of the logical error rate (LER) as the shell size increases. The simulation results for a specific device are depicted in Fig.~\ref{fig:sweet} (a) and (b). Typically, the LER for this device typically exhibits a decreasing-then-increasing trend with the growth of either the global shell size $n_\text{shell}$ or the local shell ratio $r$. Hence, there exists an optimal point for the global shell size or the local shell ratio. This trend is consistent across most random devices. The point with the lowest LER is termed the ``sweet point" for each device. In our simulation, we use LER at the ``sweet point" to represent the LER for each random device.

\begin{figure}[t]
    \centering
    \includegraphics[width=\linewidth]{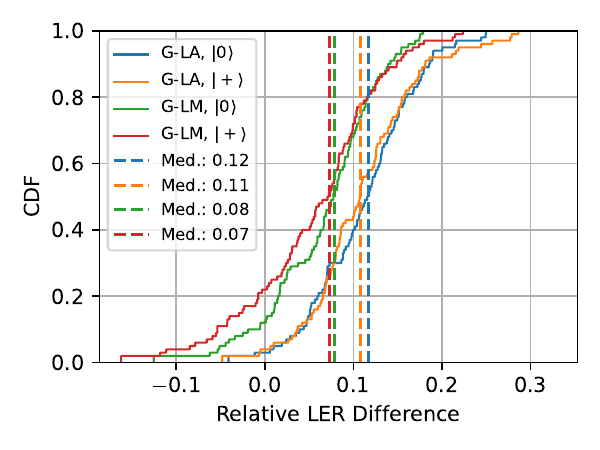}
    \caption{\textbf{Cumulative Distribution Function (CDF) shows the difference in relative LER between the GLOBAL(G), LOCALAVG(LA) and LOCALMAX(LM) Strategies.}
    G-LA represents $(\text{LER}_\text{G}-\text{LER}_\text{LA})/\text{LER}_\text{LA}$;
    G-LM represents $(\text{LER}_\text{G}-\text{LER}_\text{LM})/\text{LER}_\text{LM}$.
    The colored vertical dashed lines indicate the medians on the plots. In our simulation, we randomly generated 100 devices with $L=21$ and a defect rate of 0.02 for qubits and couplers, using a physical error rate of $p=0.002$.}
    \label{fig:gvl}
\end{figure}

We then conducted a comparison of the performance differences between the global and local shell strategies. In our simulation, we randomly generated 100 devices with code size $L=21$ and a defect rate of 0.02. We compared the relative LER difference at the sweet point to investigate the variance in error correction ability between different shell strategies, as depicted in Fig.~\ref{fig:gvl}.
From the results, we observed that the LOCALAVG strategy exhibited a greater advantage over the GLOBAL strategy compared to the LOCALMAX strategy.
The GLOBAL strategy exhibits a median 12\% increase in LER compared to the LOCALAVG strategy.
This suggests that customizing a local shell strategy for each stabilizer subgroup yields better performance than the global approach. Moreover, different local strategies demonstrate varying performance disparities. For achieving higher error correction performance, further customization of local strategies based on experimental system requirements may be warranted. However, it's worth noting that our current analysis overlooks complex noise sources such as leakage and crosstalk. When considering these noise factors, the global strategy might have advantages, particularly when calibrating a few parallel CNOT or CZ patterns. Nevertheless, we refrain from delving into detailed discussions on this aspect here. For simplicity, we primarily employ the global shell method in most of our simulations in the main text.

\section{The Advantage of Bandage-Like Super-Stabilizers}

\subsection{Advantages over Traditional Methods in Handling Bridge Syndrome Qubits (Using the Case of Fig.~3(a) in the Main Text as an Example)}

\begin{figure}[ht]
    \centering
    \includegraphics[width=\linewidth]{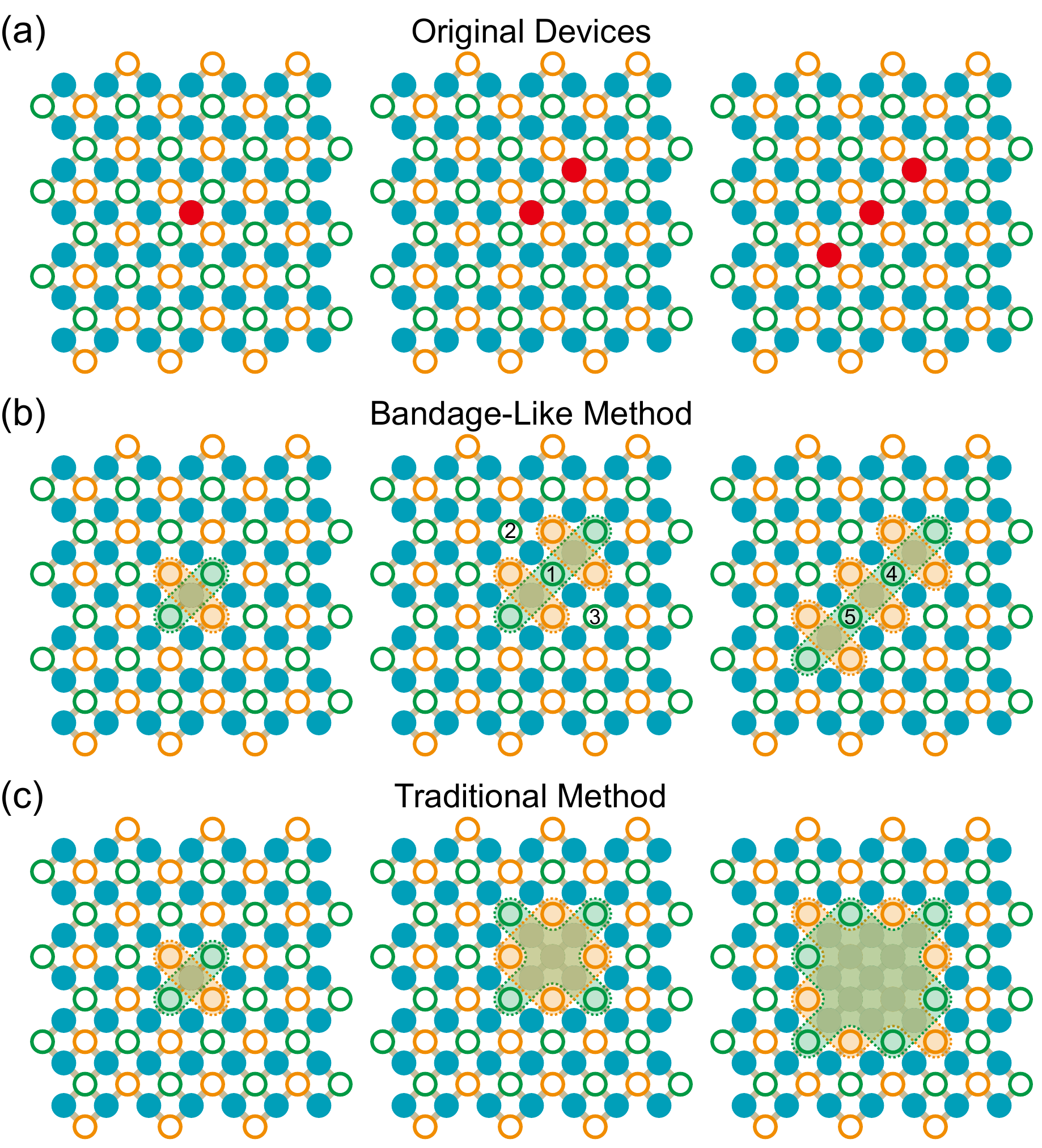}
    \caption{\textbf{An example is provided to illustrate the difference between the bandage-like and traditional methods in handling bridge syndrome qubits.} (a) A lattice with a code size of $L=7$ as mentioned in main text Fig.~3(a), with 1, 2, and 3 diagonal data qubit defects from left to right.
             (b) The lattice obtained by the bandage-like method. Only defective data qubits are disabled.
             (c)  The lattice obtained by the traditional method.
             Bridge syndrome qubits (and popped-up weight-1 syndrome qubits) are disabled, leading to the disabling of more qubits, resulting in the formation of a large area of disabled qubits protected by large weight super-stabilizers.
            }
    \label{fig:cmp123}
\end{figure}

In comparison to traditional methods, our approach retains bridge syndrome qubits instead of removing them when dealing with internal defects. This could potentially result in obtaining a larger code distance. Here, we illustrate and analyze this phenomenon using a specific example from the main text in Fig.~3(a): an $L=7$ lattice with 1 (left panel of Fig.~\ref{fig:cmp123}(a)), 2 (middle panel of Fig.~\ref{fig:cmp123}(a)), and 3 (right panel of Fig.~\ref{fig:cmp123}(a)) diagonal nearest neighbor data qubit defects.

In the case of 1 data qubit defect (left panel of Fig.~\ref{fig:cmp123}(a)), both methods show no  difference (see left panel of Fig.~\ref{fig:cmp123}(b) and Fig.~\ref{fig:cmp123}(c)), as no bridge syndrome qubits appear.

For 2 nearest neighbor data qubit defects (middle panel of Fig.~\ref{fig:cmp123}(a)), the bandage-like method demonstrates an advantage over the traditional one. In the traditional method, after removing the defective data qubits, a bridge syndrome qubit marked as $\textcircled{1}$ appears. This qubit and its surrounding data qubits need to be removed to form the super-stabilizer shown in Fig.~\ref{fig:cmp123}(c). However, with the bandage-like method, no additional qubits need to be removed after disabling the defective data qubits, enabling the surface code to function normally with bandage-like super-stabilizers. As a result, the $Z$ distance improves from 5 (traditional) to 6 (bandage-like), while the $X$ distance remains 5 for both methods. Moreover, the average super-stabilizer weight decreases from 10 (traditional) to 6.67 (bandage-like). By preserving the syndrome qubit $\textcircled{1}$, not only two additional data qubits are saved, but also two more syndrome qubits marked as $\textcircled{2}$ and $\textcircled{3}$ remain unaffected by defects, providing extra information for error correction.

In the case of 3 nearest neighbor data qubit defects (right panel of Fig.~\ref{fig:cmp123}(a)), the advantage of the bandage-like method is further amplified. Two bridge syndrome qubits marked as $\textcircled{4}$ and $\textcircled{5}$ in Fig.~\ref{fig:cmp123}(b) are preserved, leading to an improvement in $Z$ distance from 4 (traditional) to 6 (bandage-like), and a decrease in the average super-stabilizer weight from 14 (traditional) to 7 (bandage-like).

\subsection{Advantages over Traditional Methods in Handling Weight-1 Syndrome Qubits}

\begin{figure}[ht]
    \centering
    \includegraphics[width=0.7\linewidth]{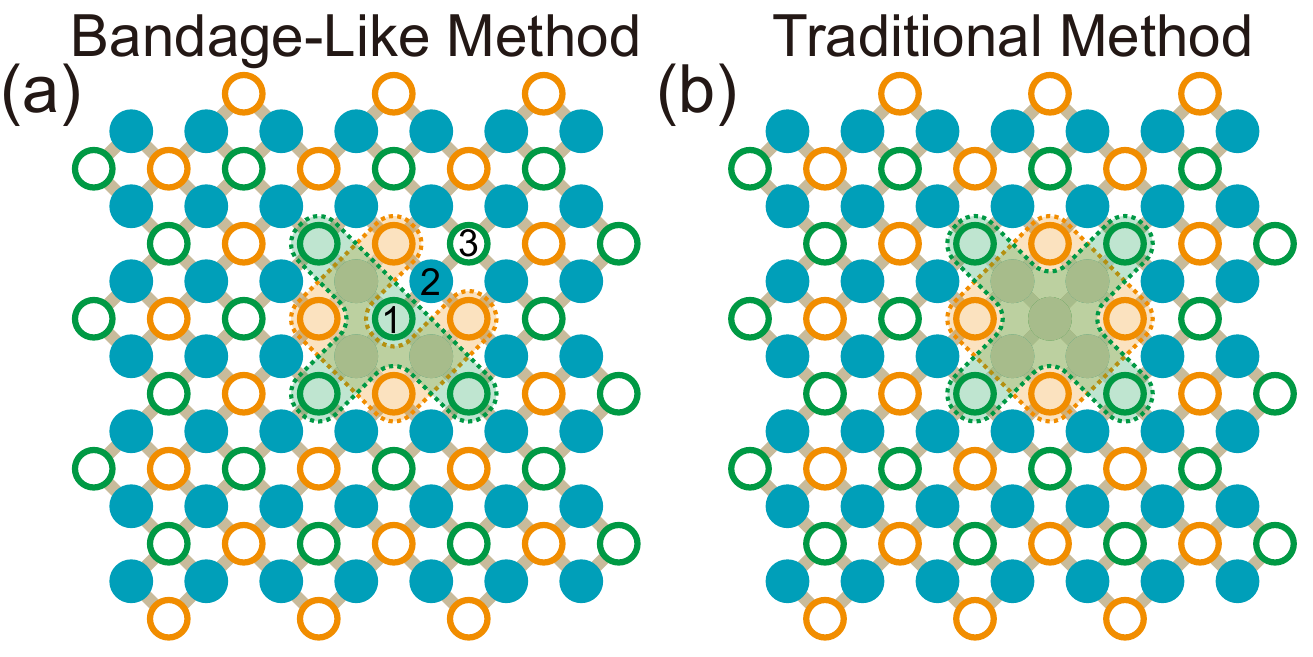}
    \caption{\textbf{An example illustrates the contrast between the bandage-like and traditional methods in handling weight-1 syndrome qubits.} (a) Bandage-like method: Depicts a scenario with a weight-1 $Z$ syndrome qubit, labeled as $\textcircled{1}$. It forms a C-shaped $X$ bandage-like super-stabilizer and a T-shaped $Z$ bandage-like super-stabilizer. By preserving weight-1 syndrome qubit $\textcircled{1}$, qubit $\textcircled{2}$ is also preserved, while syndrome qubit $\textcircled{3}$ remains unaffected by defects. (b) Traditional method: Involves disabling the weight-1 syndrome qubit and its neighboring data qubit, resulting in the formation of a higher-weight $Z$ super-stabilizer.}
    \label{fig:weight1}
\end{figure}

Our bandage-like method also preserves weight-1 syndrome qubits, unlike the traditional approach where they are disabled. Illustrated in Fig.~\ref{fig:weight1}(a), the weight-1 syndrome qubit marked as $\textcircled{1}$ is retained, resulting in a C-shaped weight-8 $X$ super-stabilizer and a T-shaped weight-10 $Z$ super-stabilizer. However, in contrast to the traditional method shown in Fig.~\ref{fig:weight1}(b), the weight-1 syndrome qubit $\textcircled{1}$ would be disabled, and further the data qubit marked as $\textcircled{2}$ must also be disabled. Consequently, our method yields a weight of 10 for the $Z$ super-stabilizer, lower than the traditional method's 12, while maintaining the $X$ super-stabilizer's weight at 8. Additionally, we gain an extra normal $Z$ syndrome qubit $\textcircled{3}$, enhancing error correction by providing more information. 

It's noteworthy that qubit $\textcircled{2}$ in Fig.~\ref{fig:weight1}(a) is included twice in the $X$ super-stabilizer, indicating that the $X$ super-stabilizer will not be triggered if a $Z$ error occurs on it. However, according to our logical operator placement strategy, no $X$ logical operator will pass through the qubit $\textcircled{2}$. Thus, disregarding $Z$ errors on qubit $\textcircled{2}$ does not negatively impact the integrity of the code.

\subsection{More Shortcomings of Traditional Methods}

Traditional methods disable weight-1 and bridge syndrome qubits when handling internal defects, which can introduce more issues, as described below.

\subsubsection{Avalanche Effect}

In the process of disabling weight-1 and bridge syndrome qubits, traditional methods may generate new weight-1 and bridge syndrome qubits. Therefore, this process needs to be iterated until no new weight-1 and bridge syndrome qubits are generated. This iterative process may require many iterations and result in the removal of many qubits, a phenomenon referred to as the avalanche effect.

\begin{figure}[h]
    \centering
    \includegraphics[width=0.7\linewidth]{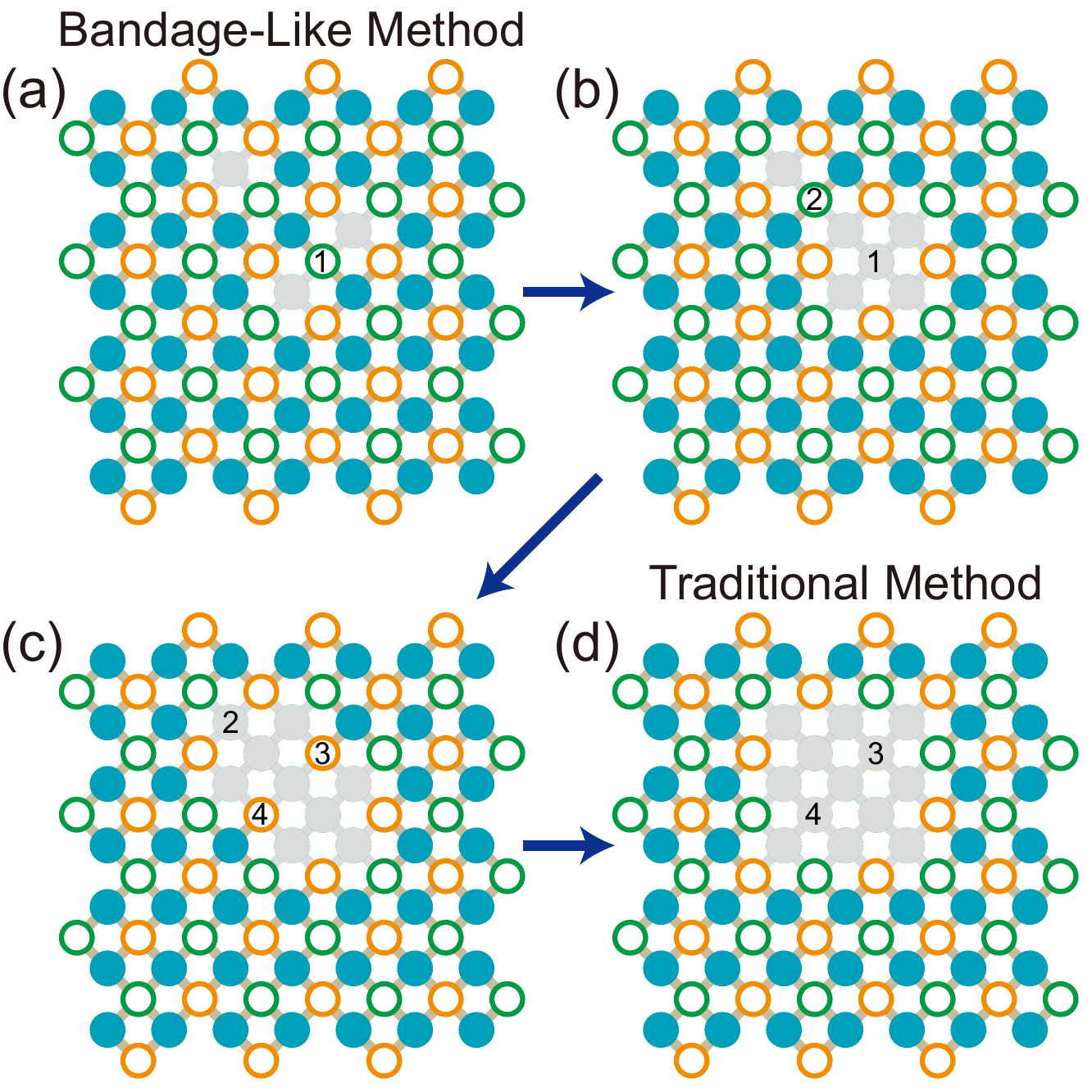}
    \caption{\textbf{An example illustrates the avalanche effect of the traditional method.}
            (a) A device generated by the internal defect-disabling rules of the bandage-like method. Three data qubits are disabled. A bridge syndrome qubit exists, marked as $\textcircled{1}$. If the traditional method is used, this bridge syndrome qubit needs to be removed.
            (b) Using the traditional method to remove bridge syndrome qubit $\textcircled{1}$ results in the emergence of a new bridge syndrome qubit, marked as $\textcircled{2}$.
            (c) Continuing with the traditional method to remove bridge syndrome qubit $\textcircled{2}$ leads to the appearance of two new weight-1 syndrome qubits, marked as $\textcircled{3}$ and $\textcircled{4}$.
            (d) The two weight-1 syndrome qubits, $\textcircled{3}$ and $\textcircled{4}$, are then disabled.
            Finally, the iterative disabling process results in a significant area of disabled qubits.}
    \label{fig:ava}
\end{figure}

Figure~\ref{fig:ava} provides a very intuitive example. For the surface code lattice in Fig.~\ref{fig:ava}(a), after three iterations of the traditional method to disable all weight-1 and bridge syndrome qubits, a total of 13 qubits are disabled. However, in the bandage-like method, only the 3 defective qubits will be disabled, maintaining the least overhead.

\subsubsection{Boundary Affected by Traditional Internal Defect Disabling Rules}

In our method, the second step of handling internal defects does not affect the boundary qubits anymore, whereas in traditional methods, disabling weight-1 and bridge syndrome qubits neighboring boundary data qubits may impact the boundary.

\begin{figure}[tbp]
    \centering
    \includegraphics[width=0.7\linewidth]{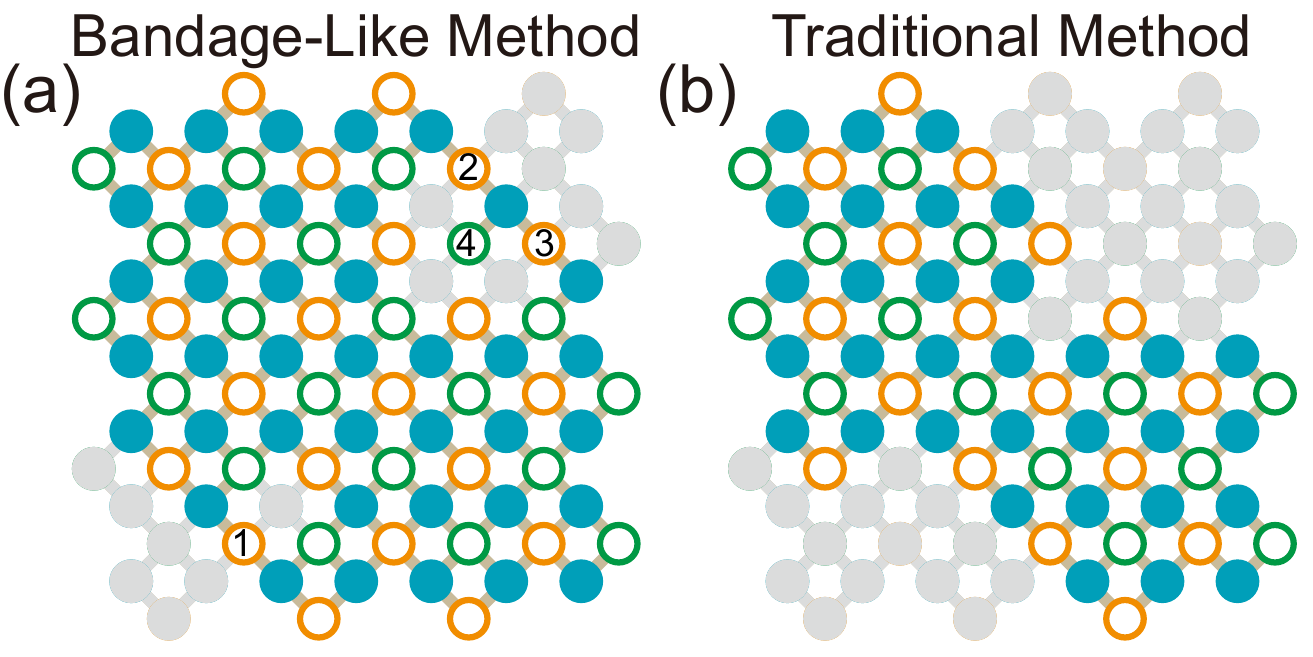}
    \caption{\textbf{An example illustrates the traditional method's potential impact on the boundary when handling internal defects.} 
    (a) The device processed using the internal defect-disabling rules of the bandage-like method. Weight-1 syndrome qubits (marked as $\textcircled{4}$) and bridge syndrome qubits (marked as $\textcircled{1}$-$\textcircled{3}$) are adjacent to boundary data qubits.
    (b) The traditional method requires further disabling of $\textcircled{1}$-$\textcircled{4}$, leading to the disabling of more qubits and altering the shape of the boundary.}
    \label{fig:bb}
\end{figure}

An intuitive example is illustrated in Fig.~\ref{fig:bb}. For the surface code lattice in Fig.~\ref{fig:bb}(a), the bandage-like method does not require further disabling of qubits, while in the traditional method, additional weight-1 and bridge syndrome qubits need to be disabled, resulting in the boundary being affected, as shown in Fig.~\ref{fig:bb}(b). One drawback of this process is the need for repeated execution of \textit{boundary deformation} and \textit{internal defect disabling} steps until no unsafe qubits remain. However, in our method, this situation does not arise, as only one iteration of boundary deformation and internal defect disabling is required.

It is worth mentioning that, for simplicity, when dealing with internal defects in both the main text and appendix, we do not remove weight-1 and bridge syndrome qubits neighboring boundary data qubits, even in traditional methods. Therefore, in practice, the modified traditional method we use as the baseline should perform better than the original traditional method.

\subsection{Extended Data for the Main Text}
\subsubsection{Extended Data for Fig.~1 in the Main Text}

\begin{figure}[t]
    \centering
    \includegraphics[width=\linewidth]{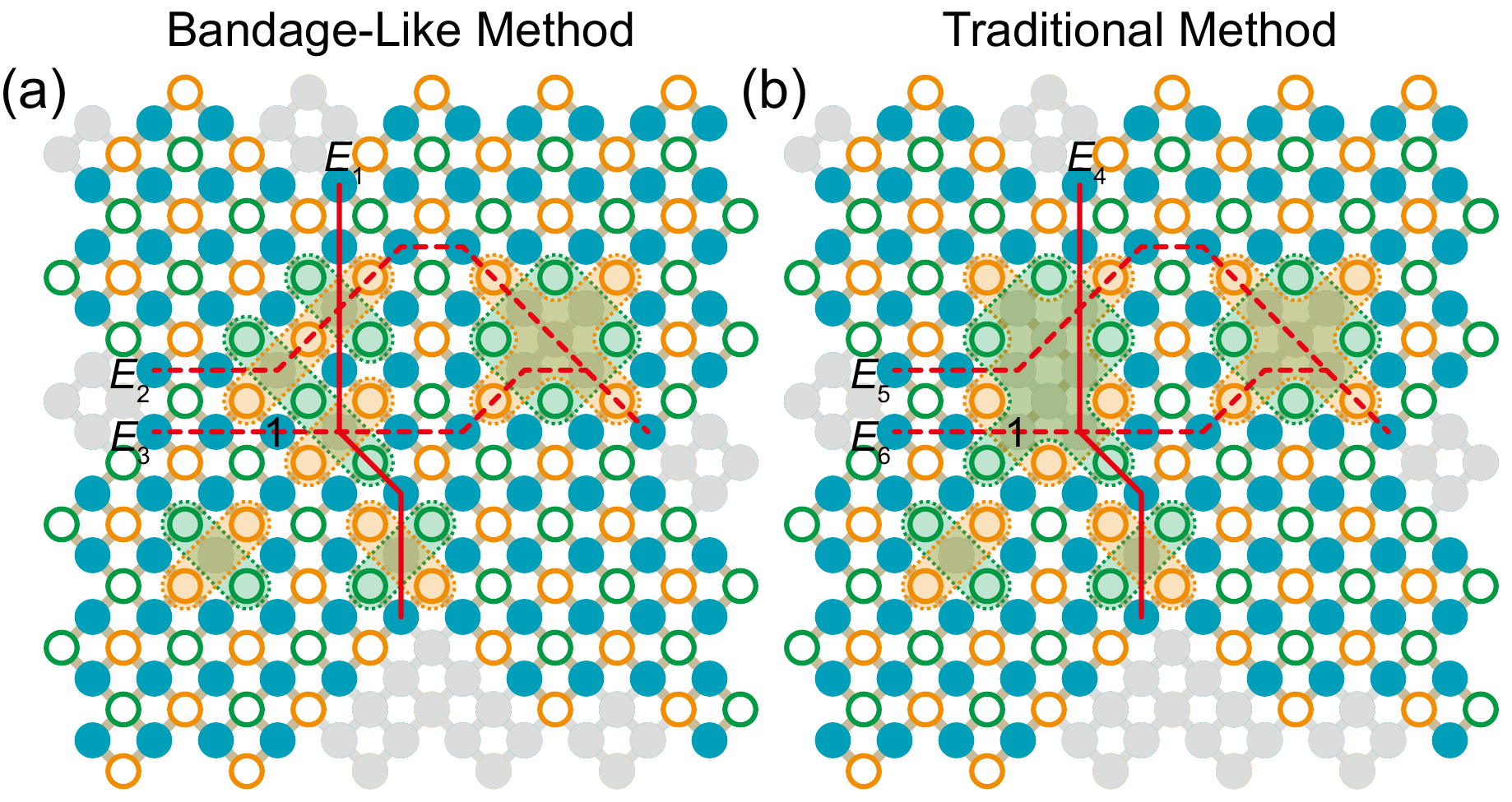}
    \caption{\textbf{Comparison between the bandage-like method and traditional method in handling the defective lattice shown in Fig.~1(a) of the main text.} (a) The result from the bandage-like method. (b) The result from the traditional method. The red lines indicate the logical error strings occurring on undisabled data qubits that trigger no stabilizer. The solid line marked as $E_1$ and $E_4$ represents the $X$ error string, while the dashed line marked as $E_2$, $E_3$, $E_5$ and $E_6$ represent $Z$ error strings.
    }
    \label{fig:admp}
\end{figure}

\begin{figure*}[t]
    \centering
    \includegraphics[width=\linewidth]{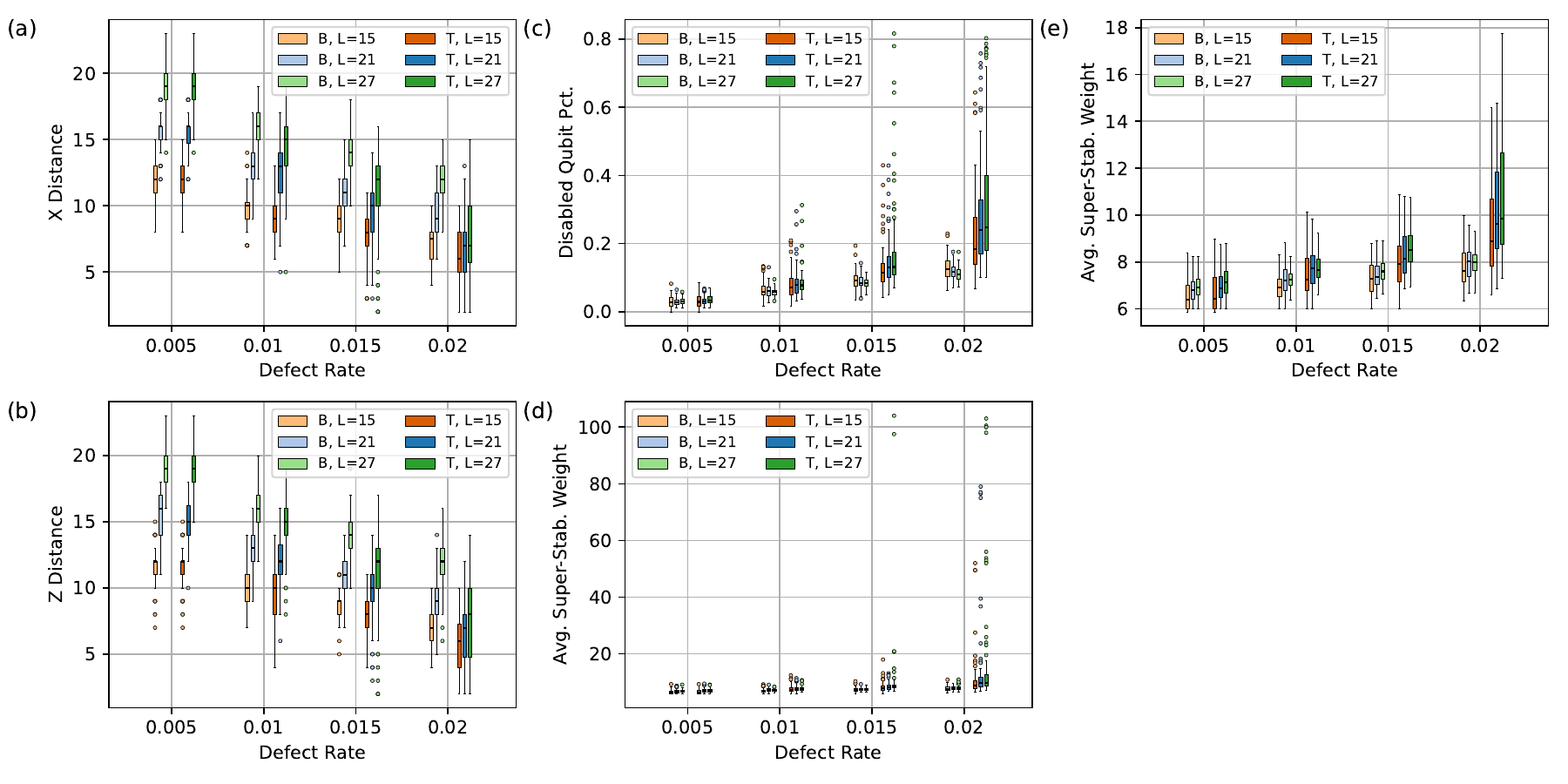}
    \caption{\textbf{Extended Data for Fig.~3(c-f) in the Main Text.} The whiskers extend to data within 1.5 times the IQR from the box. Points beyond are fliers. (a-d) display statistics for the bandage-like (B) and traditional (T) methods regarding (a) Average $X$ distance, (b) Average $Z$ distance, (c) Average disabled qubit percentage, and (d) Average super-stabilizer weight across different defect rates. (e) Average super-stabilizer weight without fliers.}
    \label{fig:basp}
\end{figure*}

For the defective lattice shown in Fig.~1(a) of the main text, we compare the surface codes obtained using the bandage-like method and traditional methods for handling defects, illustrated in Fig.~\ref{fig:admp}(a) and Fig.~\ref{fig:admp}(b) respectively.

To showcase the effectiveness of the bandage-like method in improving the code distance, we examine the shortest error strings that do not trigger any stabilizer. In Fig.~\ref{fig:admp}(a), using the bandage-like method, we find that the shortest $X$ error string labeled as $E_1$ consists of 5 undisabled data qubits, with a weight of 5. Consequently, the $X$ distance achieved with the bandage-like method is 5. Conversely, in Fig.~\ref{fig:admp}(b), employing the traditional method, the shortest $X$ error string labeled as $E_4$ consists of 4 undisabled data qubits, with a weight of 4. Thus, the $X$ distance of the code generated by the bandage-like method exceeds that of the traditional method. Additionally, the bandage-like method reduces the average super-stabilizer weight from 8.5 to 7.2 and disables 5 fewer qubits.

For $Z$ errors, both the shortest $Z$ logical error strings marked as $E_2$ and $E_5$ have a weight of 5. Therefore, the $Z$ distance of the bandage-like method is not improved.
However, even though the $Z$ distance is not improved, saving more qubits in the bandage-like method increases the weight of some logical error strings.
For instance, consider a data qubit marked as $\textcircled{1}$ saved in the bandage-like method. Qubit $\textcircled{1}$ elevates a weight-5 $Z$ error string marked as $E_6$ in the traditional method to a weight-6 string marked as $E_3$ in the bandage-like method.
In this example device, 12 weight-5 $Z$ logical error strings are elevated, reducing the total number of weight-5 $Z$ error strings from 18 (traditional) to 6 (bandage-like).
Elevating the weight of logical error strings led to a decrease in their probability of occurrence, ultimately reducing the logical error rate. This phenomenon is also discussed in \cite{lin2024codesign}, which highlights the correlation between error correction ability and the number of unique weight-$d$ logical operators.

\begin{figure}[t]
    \centering
    \includegraphics[width=\linewidth]{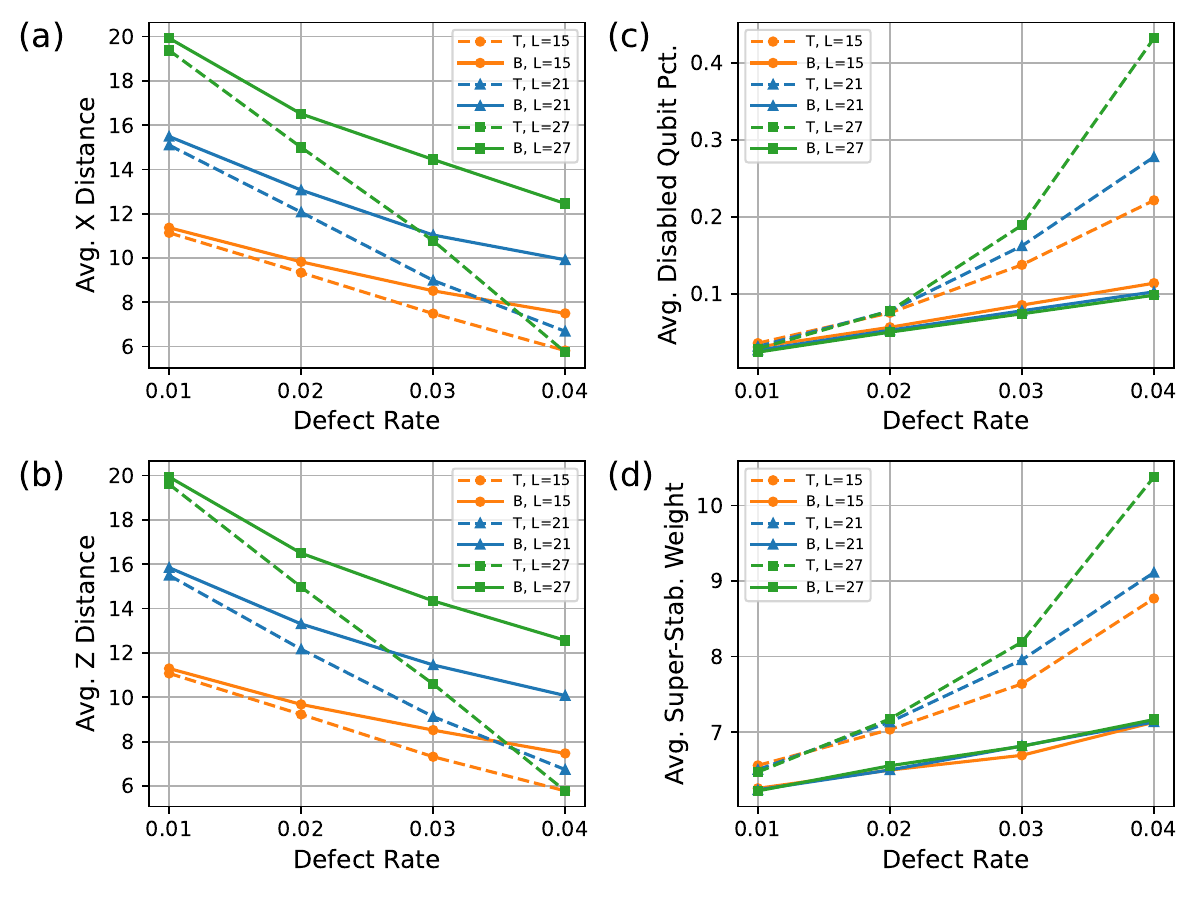}
    \caption{\textbf{Performance comparison of the traditional method with our approach in handling defects for the scenarios of devices with only coupler defects} (a-d) display statistics for the bandage-like (B) and traditional (T) methods regarding (a) Average $X$ distance, (v) Average $Z$ distance, (c) Average disabled qubit percentage, and (d) Average super-stabilizer weight across different defect rates. The defects occur exclusively in couplers.}
    \label{fig:baco}
\end{figure}

\subsubsection{Extended Data for Fig.~3(c-f) in the Main Text}

Here, we provide comprehensive boxplots illustrating the advantages in terms of code distance preservation, disabled qubit percentage, and average super-stabilizer weight, as discussed in Fig.~3(c-f) of the main text. The results are depicted in Fig.~\ref{fig:basp}. For each defect rate and code size $L$, we generated 100 devices with randomly distributed
defects. Notably, the results for disabled qubit percentage and average super-stabilizer weight display numerous fliers for the traditional method at higher defect rates. This indicates that as the defect rate increases, defects tend to cluster, resulting in a significant overhead for the traditional method in handling such situations. Occasionally, the traditional method for adapting devices fails to generate a correct stabilizer measurement circuit at high error rates due to excessive qubit disabling. Conversely, the bandage-like method adeptly manages such situations and successfully generates stabilizer measurement circuits for all cases in the simulation. For better clarity, the average super-stabilizer weight without fliers is presented in Fig.~\ref{fig:basp}(e).

\subsubsection{Scenarios of Devices with Only Coupler Defects}

In the main text, our discussion mainly focuses on scenarios where qubits and couplers share the same defect rate. In fixed-frequency transmon setups, frequency collisions are the most significant fabrication defects, treated as coupler defects in our method. Continuous coupler defects often result in weight-1 syndrome qubits and bridge syndrome qubits, leading to the traditional method disabling more qubits and resulting in higher-weight super-stabilizers. However, the bandage-like method effectively mitigates this issue with minimal additional costs. Therefore, in scenarios with only coupler defects, the bandage-like method demonstrates even more significant advantages over the traditional approach.

We considered extreme scenarios of devices with only coupler defects. Similar to Fig.~3(c-f) in the main text, we simulated the variations of code distance, disabled qubit percentage, and average super-stabilizer weight with changing defect rates and code size $L$, as shown in Fig.~\ref{fig:baco}. It's evident that under conditions where only coupler defects are present, the bandage-like method exhibits even greater advantages compared to traditional methods. For instance, at a code size of $L=27$ and a defect rate of 0.04 for the couplers
(the defect rate on the entire lattice is greater than the situation where both coupler and qubit defect rates are 0.02),
the average $X$ distance improves from 5.8 to 12.5, and the $Z$ distance from 5.8 to 12.6, representing a 117\% average improvement. Meanwhile, the disabled qubit percentage decreases from 43.2\% to 9.8\%, and the average super-stabilizer weight decreases from 10.4 to 7.2, marking a 31\% improvement. This increase in improvement is significantly higher than scenarios where qubits and couplers share the same defect rate as shown in the main text.

\bibliographystyle{apsrev4-1}
\bibliography{references}